# Cascade of Phase Transitions and Dirac Revivals in Magic Angle Graphene

U. Zondiner[1†], A. Rozen[1†], D. Rodan-Legrain[2†], Y. Cao[2], R. Queiroz[1], T. Taniguchi[3], K. Watanabe[3], Y. Oreg[1], F. von Oppen[4], Ady Stern[1], E. Berg[1], P. Jarillo-Herrero[2*] and S. Ilani[1*]

[1] *Department of Condensed Matter Physics, Weizmann Institute of Science, Rehovot 76100, Israel.*
[2] *Department of Physics, Massachusetts Institute of Technology, Cambridge, Massachusetts 02139, USA.*
[3] *National Institute for Materials Science, 1-1 Namiki, Tsukuba, 305-0044 Japan.*
[4] *Dahlem Center for Complex Quantum Systems and Fachbereich Physik, Freie Universität Berlin, 14195 Berlin, Germany.*
[†] These authors contributed equally to the work.
[*]Correspondence to: pjarillo@mit.edu and shahal.ilani@weizmann.ac.il

**Twisted bilayer graphene near the magic angle exhibits remarkably rich electron correlation physics, displaying insulating, magnetic, and superconducting phases. Here, using measurements of the local electronic compressibility, we reveal that these phases originate from a high-energy state with an unusual sequence of band populations. As carriers are added to the system, rather than filling all the four spin and valley flavors equally, we find that the population occurs through a sequence of sharp phase transitions, which appear as strong asymmetric jumps of the electronic compressibility near integer fillings of the moiré lattice. At each transition, a single spin/valley flavor takes all the carriers from its partially filled peers, "resetting" them back to the vicinity of the charge neutrality point. As a result, the Dirac-like character observed near the charge neutrality reappears after each integer filling.** Measurement of the in-plane magnetic field dependence of the chemical potential near filling factor one reveals a large spontaneous magnetization, further substantiating this picture of a cascade of symmetry breakings. The sequence of phase transitions and Dirac revivals is observed at temperatures well above the onset of the superconducting and correlated insulating states. This indicates that the state we reveal here, with its strongly broken electronic flavor symmetry and revived Dirac-like electronic character, is a key player in the physics of magic angle graphene, forming the parent state out of which the more fragile superconducting and correlated insulating ground states emerge.

When a system of interacting electrons is cooled from high to low temperatures, it gradually loses its high energy degrees of freedom while developing correlations that give rise to new collective degrees of freedom. The result is a hierarchy of energy scales, where the effective degrees of freedom at each scale are built out of a subset of those of the preceding ones. Understanding the nature of the effective degrees of freedom and the correlations among them at higher energies is therefore foundational for understanding the more fragile states that emerge at lower energies.

The recent discovery that twisted bilayer graphene (TBG) exhibits correlated insulating[1], superconducting[2,3], and ferromagnetic[4,5] phases has opened a fascinating playground to study such strongly correlated states. For twist angles that approach a 'magic angle' (MA) of $\sim 1.1°$, the bands in this system were predicted[6] to narrow down dramatically[7–9], leading to a variety of possible symmetry breaking ground states[10–16]. These bands, identified directly by spectroscopic[17–19] and capacitive measurements[20] were shown to significantly broaden due to interactions[21–23] and exhibited correlation-induced gaps at the Fermi energy[21,22,24] as well as lattice symmetry breaking[21,22,24]. Yet, a key question remains: what is the nature of the correlated state at energy scales comparable to the bandwidth, which gives rise to all the low energy many-body ground states.

In this work, using scanning measurements of the local electronic compressibility, we reveal that the energy bands of MATBG undergo a dramatic transformation near integer fillings. Starting at the charge neutrality point (CNP), the four spin/valley flavors exhibit a Dirac-like character. However, as integer fillings are approached, sharp transitions occur in which a single flavor takes all the carriers from its partially filled peers, "resetting" them back to the vicinity of the CNP. This Sisyphean band filling, where the electrons attempt to fill all flavors symmetrically, yet repeatedly slip back to the Dirac point for some of the flavors, revives the Dirac-like character of electrons at the Fermi level at integer fillings. Near these transitions, the compressibility reveals a striking asymmetry of the electronic state: the compressibility is high below any of the transitions, and drops precipitously above them. This behavior is observed at temperatures well above the onset of the correlated insulator and superconducting states, demonstrating that the state we reveal here, with its strongly broken electronic flavor symmetry and revived Dirac-like electronic character is a key driving force in the physics of MATBG, and should serve as a starting point for understanding the more fragile superconducting and correlated insulating ground states.

Our device consists of boron nitride-encapsulated TBG, fabricated by the "tear-and-stack" method[25,26], and placed over a local metallic back-gate[1,2] (fig. 1a). It features a large ($5\mu m \times 5\mu m$) area for imaging compressibility, and multiple contacts for transport characterization. The results reported in the main paper are based on a systematic study of a large region in the sample. We observe similar behavior in spatially separated and completely independent parts of the sample, confirming that our observations are generic (Supp. Info. SI6). Figure 1b shows the four probe longitudinal resistance as a function of carrier density, $n$ (controlled by the back gate voltage, $V_{BG}$) and parallel magnetic field, $B_{||}$, at a temperature of $T = 50mK$. These measurements exhibit insulating peaks when the flat bands are completely full or empty, $\nu = n/(n_s/4) = \pm 4$ ($n_s$ is the density corresponding to four electrons per moiré unit cell), and at half filling, $\nu = \pm 2$. Incipient superconductivity is reflected in resistance drops observed on the electron-doped side of $\nu = 2$ (up to $B_{||} \approx 0.5T$) and hole-doped side of $\nu = -2$ (up to $B_{||} \approx 1T$) similar to the phenomenology observed in devices near the MA[2,3]. The finite resistance of both the insulating and superconducting states in our case indicates that magic angle patches appear in series/parallel to non-magic-angle regions. Transport measurements through various contact pairs indeed suggest spatial variability of $n_s$ across the device, reflecting ~10% (0.1°) modulations of the twist angle (SI5). This will be confirmed more clearly by the local imaging measurements below.

To image the inverse compressibility we use a nanotube-based scanning single electron transistor[27,28] (SET). We keep the TBG grounded and monitor with the SET the local electrostatic potential modulations above the TBG in response to modulations of the back-gate voltage. This yields the local inverse compressibility, $d\mu/dn$ (SI2). Figure 1c shows a typical trace of $d\mu/dn$ vs. $n$, measured at $T = 4K$ at the center of the sample, where the angle is smaller than the MA. The increased $d\mu/dn$ around the CNP reflects the low density of states (DOS) of the Dirac-like bands near this filling. The two sharp $d\mu/dn$ peaks at $\nu = \pm 4$ originate from the energy gaps between the flat bands and the higher energy dispersive bands. Integrating this curve yields the chemical potential, $\mu$, vs. $n$ (blue fig. 1c). In this curve the energy gaps are apparent as jumps in $\mu$, and a Dirac-like behavior of the carriers shows up as a $\sqrt{n}$-like dependence of $\mu$ around the CNP.

We characterize the variations of the twist angle across the sample[29,30] by tracking the spatial variation of back-gate voltage, $V_{BG}(r)$, at which the $\nu = 4$ inverse compressibility peak appears. From this we deduce $n_S(r) = \epsilon V_{BG}(r)/d$, where $d$ is the AFM measured distance to the back gate and $\epsilon = 3.3 \pm 0.5$ is the dielectric constant of h-BN. The local twist angle then follows from $\theta(r) = a\sqrt{n_s(r)\sqrt{3}/8}$, where $a = 0.246nm$ is the graphene lattice constant. Figure 1d shows $d\mu/dn$ measured as a function of the spatial position, $Y$, along a linecut across the sample (dashed red) and as a function of $V_{BG}$. Visibly, the $\nu = 4$ peak shifts from $V_{BG} = 5.4V$ at the sample center to $V_{BG} = 6.5V$ near the top edge, corresponding to ~10% change in $\theta$ (top axis). Notably, this variation occurs in discrete steps, reflecting domains with constant $\theta$ within the sample[30]. Indeed, mapping $\theta$ across the entire accessible area (fig. 1e) shows a terraced $\theta$ landscape, with constant-$\theta$ domains of a few hundred $nm$ in size. We will start our study at the center of the sample, where $\theta$ is smaller than the MA but has minimal spatial variation (red arrow, fig. 1d) and then go to the top edge of the sample, where $\theta$ approaches the MA (cyan arrow, fig. 1d).

Striking features emerge in the electronic compressibility already for $\theta$ smaller than the MA. Figure 2a shows two $d\mu/dn(\nu)$ curves, measured at the center of the sample ($\theta = 0.99°$), at $T = 4K$ and $50mK$ in zero in-plane magnetic field, $B_\parallel = 0T$. At the lower temperature, $d\mu/dn$ exhibits a sharp and asymmetric feature near $\nu = 2$ (black arrow): approaching $\nu = 2$ from below, $d\mu/dn$ drops rapidly to zero (corresponding to infinite compressibility), jumps up sharply near $\nu = 2$ (to low compressibility), and then decreases gradually with further increase of $\nu$. This asymmetric sawtooth-like feature in $d\mu/dn$ is fundamentally different from what is expected for a gap opening: A gap is associated with a jump in chemical potential, $\mu$, and thus corresponds to a narrow *peak* in $d\mu/dn$ which is symmetric around the transition (even when the gap is driven by interactions, e.g., as seen in fractional quantum Hall gaps[31]). The absence of such a peak in our measurement thus puts a tight upper bound on the size of a thermodynamic gap in this system (SI7).

The sawtooth features in $d\mu/dn$ become even more dramatic when $\theta$ approaches the MA. We start by studying the $\theta$ dependence of the compressibility at finite $B_{||}$, which strengthens the features at quarter fillings ($\nu = 1,3$) and then proceed to explore the physics at $B_{||} = 0$. Figure 2b plots a color map of $d\mu/dn$, measured at $B_{||} = 12T$ as a function of $\nu$ and the spatial coordinate, $Y$, along the same line cut as in fig. 1d. Along this line cut, $\theta$ gradually climbs toward the MA (right y-axis). We define the $\nu$ axis by separately normalizing the density at each $Y$ with respect to the local moiré band density, $\nu = n/(n_s(Y)/4)$. Figure 2c exhibits selected curves from this spatial evolution, offset for clarity. Interestingly, we now observe a cascade of sawtooth features, which appear at almost all integer fillings, both within the conduction and the valence flat bands. The jumps in $d\mu/dn$ occur close to the integer values or slightly above them, are often quite abrupt, and are consistently followed by a gradual decrease of $d\mu/dn$ with $n$. Remarkably, all the observed sawtooth features are pointing away from the CNP. Moreover, while there is an approximate mirror symmetry of the sawtooth directions about the CNP, there are clear *asymmetries* about the centers of the conduction and valence flat bands ($\nu = 2, -2$): $d\mu/dn$ at $|\nu| = 2 + x$ is radically different from $d\mu/dn$ at $|\nu| = 2 - x$. These observations emphasize the unique role of the CNP in the physics, and strongly indicate that an inherent asymmetry of the single-particle dispersion around the center of the conduction and valence bands must be instrumental for the observed asymmetry in the many-body ground state.

Further hints for the underlying physics are provided by the high value of $d\mu/dn$ immediately after the transitions and its subsequent gradual decrease with $n$. These $d\mu/dn$ tails resemble the behavior of $d\mu/dn$ around the CNP, which suggests a revival of the Dirac electrons near integer fillings. This can also be seen by comparing measurements of $\mu(n)$ at two different twist angles (fig. 2d). At the smaller $\theta$, $\mu(n)$ has a $\sqrt{n}$-like dependence around the CNP, characteristic of a Dirac-like dispersion, but then climbs roughly linearly as the flat band is filled. In contrast, when $\theta$ is close to MA, we observe that a $\sqrt{n}$-like dependence re-emerge at every integer filling.

While at $\nu = 2, 3$ sawtooth features appear already at small $\theta$, a detailed look at the evolution of $d\mu/dn$ with twist angle reveals that when approaching the MA, the sawtooth features both increase in magnitude and move toward exact integer fillings. In contrast, the $\nu = 1, -1, -2$ sawtooth features emerge only at a larger $\theta$ and become very prominent as $\theta$ approaches the MA. We note that the sawtooth features in the valence flat band ($\nu = -1, -2$) appear only at the largest angle in our measurement, suggesting that holes need larger $\theta$ than electrons to exhibit this behavior.

A recurring feature in our measurements are dips in $d\mu/dn$ appearing in a narrow density region just prior to the sharp sawtooth jumps. Several examples are shown in the zoom-ins in fig. 2e. At these dips, $d\mu/dn$ is consistently seen to drop to zero or to a slightly negative value. We note that these dips corresponds to high compressibility, which is the opposite of what is expected from a gapped high resistance state.

The temperature dependence of $d\mu/dn$ is plotted in fig. 3a. We observe that the sawtooth features remain strong even at the largest achievable temperature in our scanning setup, $T = 16K$, and are estimated to decay to zero at $T \approx 30K$ (SI12), a factor seven higher than the typical temperatures below which insulating behavior commences in transport measurements[1,3,32–34]. Independently, we can estimate the energy scale associated with the Dirac revivals from the measured depth of the kinks in $\mu$ (fig. 2d), obtaining $\Delta\mu \sim 4mV$. This energy scale is significantly larger than the activation energies measured in transport[1,3,32], which were associated with correlated gaps, and of comparable scale to the range in which anomalous behavior was observed in transport[33,34]. These observations suggest that the observed Dirac revivals underlie the high-energy correlated state of this system.

Figure 3b shows the $B_\parallel$ dependence of $d\mu/dn$ for $\theta = 1.05°$, showing the evolution of near-MA compressibility from $B_\parallel = 12T$ to $B_\parallel = 0$. We observe that the magnitude of the sawtooth at fillings $\nu = 1, 3$ increases with $B_\parallel$, while it decreases at $\nu = -2$ and weakly decreases at $\nu = 2$. At the CNP, $d\mu/dn$ hardly changes with $B_\parallel$. This filling factor dependence reflects alternating ground state spin polarizations as integer fillings are crossed, consistent with previous transport observations[1–3,32]. Near $\nu = 2, 3$ smaller superimposed sawtooth-like features are visible (gray lines), arising from nearby domains with smaller twist angles that are also detected by the SET due to its finite spatial resolution. In contrast, at $\nu = -2, 1$ we observe a single sawtooth feature. This is consistent with our previous observation that sawtooth features emerge for a range of twist angles for $\nu = 2, 3$ but only close to the MA for $\nu = -2, 1$. Importantly, these $B_\parallel$ dependence measurements show that the compressibility near MA exhibits sawtooth features at the fractions $\nu = -2, 1, 2, 3$ even in the limit of $B_\parallel = 0$.

The prominent magnetic field dependence of the sawtooth feature around $\nu = 1$ allows us to study its energetics more carefully. Figure 3c shows the chemical potential difference from the CNP, $\mu - \mu_{CNP}$, as a function of $n$ for various $B_\parallel$. At every density, $\mu$ depends approximately linearly on $B_\parallel$, but with a different slope (the top inset shows few representative $\mu$ vs. $B_\parallel$ curves). The slope of $\mu$ with respect to $B_\parallel$ is related by a Maxwell relation to the *differential magnetization*, $dM_\parallel/dn \equiv -d\mu/dB_\parallel$, thus allowing us to determine $dM_\parallel/dn$ directly. The bottom inset of fig. 3c plots $dM_\parallel/dn$ as a function of $\nu$ for $\theta = 1.05°$ and $0.99°$. The differential magnetization at $\theta = 1.05°$ (red trace) is nearly zero over a finite filling factor range around the CNP. It then rises slowly with $\nu$, and near $\nu = 1$ exhibits a rapid increase to a value of $dM_\parallel/dn \approx 4.5 \pm 0.3\mu_B$, followed by a sharp drop to near zero. These large values of $dM_\parallel/dn$ indicate that close to $\nu=1$, magnetization builds up rapidly with few added carriers. The linear $B_\parallel$ dependence of $\mu$, observed to the lowest $B_\parallel$ in our measurements (top inset), suggests that the system has a non-zero spontaneous magnetization near and above $\nu = 1$. In contrast, the differential magnetization measured at $\theta = 0.99°$ remains practically zero throughout the same filling factor range (the full $B_\parallel$ dependence data for $\theta = 0.99°$ is shown in SI8). The difference between these two measurements highlights that the emergence of flavor symmetry breaking and magnetization near $\nu$ =1 depends sharply on the closeness of $\theta$ to the MA.

These salient, robust features of the compressibility call for a theoretical understanding. The electronic structure of the flat moiré bands in TBG is highly complex. However, as we shall now show, many features of our data are captured surprisingly well within a simple model that ignores many of these details, provided that certain key elements are included. A typical calculated band structure of the flat bands in TBG near the MA is shown in fig. 4a. Among the many details, a key feature is the strong asymmetry of conduction and valence flat bands with respect to their centers, which evolve from Dirac-like DOS at the CNP to massive DOS near the valence and conduction band edges (sometimes followed by a low DOS "tail").

We believe that two important features of this band structure are responsible for the behavior observed in our experiment. The first is the strong dependence of the DOS on filling. The second is the inherent asymmetry of this gradually increasing DOS with respect to the center of the conduction (or valance) band. To capture these features, we consider a stripped-down model where the DOS is replaced with a linear function that terminates abruptly at the flat band top and bottom (fig. 4a, right). We assume a contact repulsive interaction whose magnitude is independent of the spin/valley flavor index. Finally, the model is solved within the Hartree-Fock approximation, allowing for breaking of the spin/valley symmetry by an unequal population of electrons of different flavors. For a full description of the calculation, see SI9.

A typical result for $d\mu/dn$ for a system with an interaction strength comparable to the band width is shown in fig. 4b. The resemblance to the experimental data is apparent, with asymmetric sawtooth-like features appearing at integer filling factors, preceded by sharp drops of $d\mu/dn$ to zero. Figure 4c shows the corresponding $\mu$, which also captures the main features observed in the experiment (fig. 2d). Figure 4d shows the population of the individual flavors as a function of the total density, revealing a cascade of phase transitions upon increasing $n$. In the specific example shown in the figure we observe first-order transitions near $\nu = 1$ and $\nu = 2$, where one flavor becomes nearly filled, while the population of the other flavors is pushed downwards towards charge neutrality. Closely after these transitions, there are "Lifshitz" transitions where the filling of the majority flavor reaches 1. This Lifshitz transition explains the sawtooth features in $d\mu/dn$: before the transition, the density of states at the Fermi level is dominated by the massive majority flavor, whereas after the transition it is strongly decreased due to the Dirac character of the remaining flavors whose occupation is reset to near the CNP. Depending on the model details, the first order and Lifshitz transitions may coincide, i.e., the filling factor of the majority flavor may jump directly to 1 (SI11). We also see that flavor polarization can be initiated by a second order phase transition (e.g. preceding $\nu = 3$ in figure 4d). In our model, the sharp dips where $d\mu/dn$ goes to zero appear at the first order phase transition and are a manifestation of a spatial breakup of the system into domains of the two corresponding phases. However, we note that such $d\mu/dn \approx 0$ dips may also arise from other reasons, e.g., it may reflect a van-Hove singularity very close to the band edge.

In the SI, we elaborate on the effects of additional details not included in this minimal model, such as a more complex DOS that includes van Hove singularities (SI12) and long-range Coulomb interactions (SI15) that prevent a macroscopic phase separation at a first order transition (stabilizing instead a mesoscopic mixture of domains of the two phases). While many of the details of the computed $d\mu/dn$ change, the overall features are found to be similar to those obtained from the simple model.

Our experiments and corresponding theory predict that the flavor symmetry will be broken with a 4,3,2,1-fold degeneracy following fillings of $\nu = 0,1,2,3$ (SI16). This is indeed what was observed experimentally[1–3,30,32] (with one possible exception, above $\nu = 1$, which may be due to the rather sparse and inconclusive data there). It also predicts that Landau fans will always point away from the CNP[10], apart from the fan emerging from $\nu = 4$ that should point toward the CNP (SI16), which also reproduces the experimental observations[1–3,30,32]. Another immediate consequence of the Dirac revival physics is that correlated gaps will appear at integer filling only if the Dirac spectrum is gapped[10,14,16], suggesting a possible connection between a gap at the CNP and the correlated gaps at integer filling factors. In fact, in our data there is no thermodynamic indication of a gap, and we can place an upper bound of $\sim 0.5 meV$ on a gap that may be hidden within our experimental resolution. This is consistent with the $0.14 - 0.9 meV$ activation gaps measured so far in transport[1–3,32], but also shows that correlated-insulator physics, at least in these samples, occurs only at much lower energies than the Dirac revivals physics which survives to scales of $\sim 4 meV$. Finally, we note the resemblance between our thermodynamic measurements of $d\mu/dn$ vs. $\nu$ and $B_{\parallel}$ near $\nu = 1$ (colormap in the inset in fig. 3b) and the observation of hysteretic ferromagnetic behavior in transport[4] near $\nu = -1$ (fig. 4 in ref. [32]). This hints that the interpretation of the $d\mu/dn$ dips as markers for possible real-space phase mixing is plausible, suggesting that our observation of such dips near other filling factors implies that phase mixing regions are a rather generic feature of the physics near various integer fillings.

In summary, we have observed a sequence of sharp asymmetric jumps in the local electronic compressibility near integer filling factors in twisted bilayer graphene close to the magic angle. These are naturally interpreted in terms of breaking of the spin/valley symmetries at these fillings, which revive the Dirac-like character of the carriers after each transition. These features appear already at temperatures far above the onset of the superconducting and correlated insulating states, indicating that the state observed here is the high energy correlated state from which superconductors and insulators emerge at lower temperatures.

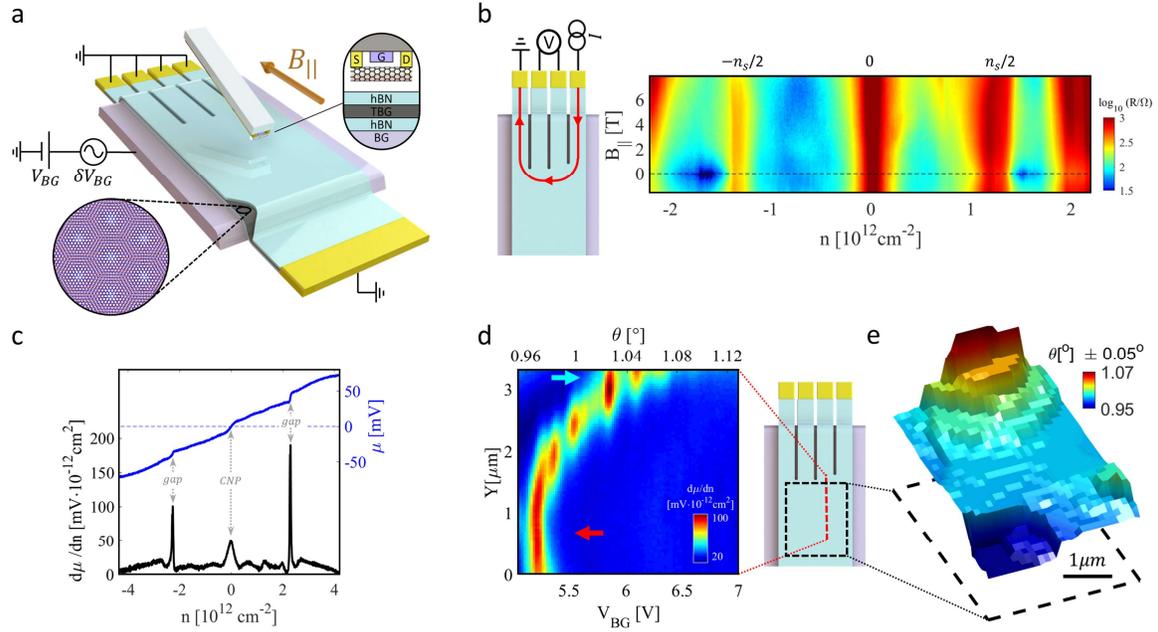

**Figure 1: Measurement setup and device characterization**. **a**. Twisted bilayer graphene (TBG), encapsulated between top and bottom h-BN (blue), placed on a metallic back-gate (purple), and contacted by side contacts (yellow). We image the inverse electronic compressibility, $d\mu/dn$, using a scanning nanotube-based single electron transistor (SET) (inset) (where $\mu$ is the chemical potential of the carriers in the TBG and $n$ is their density). The SET measures the local $\delta\mu$ in response to a density modulation, $\delta n$, produced by an AC voltage on the back-gate, $\delta V_{BG}$ (see SI2). In these measurements the TBG is kept grounded, and a DC back-gate voltage, $V_{BG}$, sets the overall $n$. Some measurements use a parallel magnetic field, $B_{\parallel}$, whose orientation is indicated by the orange arrow. **b.** Four probe resistance, $R$, measured using the top four contacts (left inset) as a function of carrier density, $n$, and $B_{\parallel}$, at $T = 50mK$. The $\pm n_s/2$ labels on the top axis mark half-filled valence and conduction flat bands. **c.** Characteristic $d\mu/dn$ (black) and $\mu$ (blue, obtained by integrating the former) measured as function of $n$, at $T = 4K$. Arrows mark the charge neutrality point (CNP) and the gaps separating the flat bands from the higher-energy dispersive bands. **d.** Measurement of $d\mu/dn$ as a function of a spatial coordinate, $Y$, along a linecut across the sample (red dashed line, inset) and $V_{BG}$, focusing on the peak at of $d\mu/dn$ that corresponds to a full flat band ($n = n_s$). The local twist angle, $\theta$, is obtained directly from the back-gate voltage at which this peak appears (see text), and is indicated in the top y-axis. In the center of the sample (red arrow), $\theta$ is rather homogenous and smaller than the magic angle. In the top part of the sample $\theta$ climbs in steps (fixed $\theta$ domains) towards the MA. **e.** Spatial map of the twist angle, $\theta$ (covering the black dashed region in the inset), determined in a similar fashion to panel d but from a three dimensional measurement of $d\mu/dn$ as a function of $V_{BG}$, $X$ and $Y$ (SI4). A terraced landscape with constant-$\theta$ domains is visible.

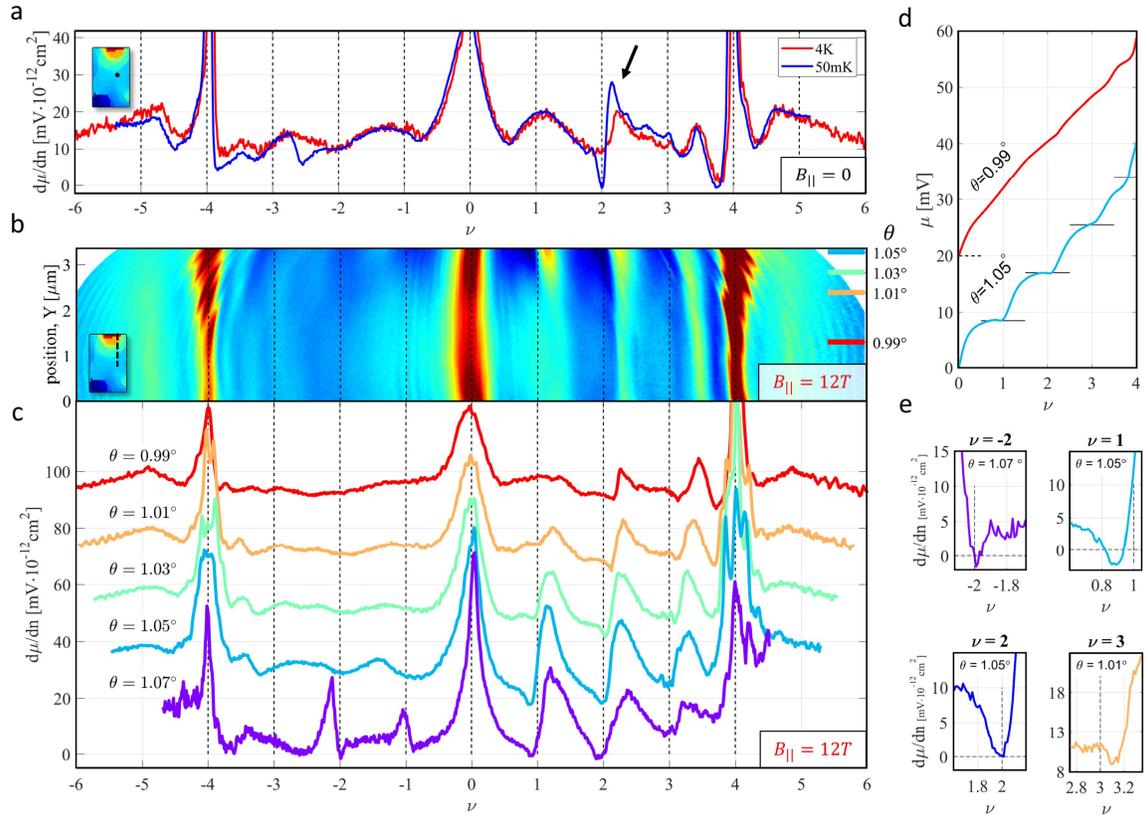

**Figure 2: Asymmetric sawtooth features in inverse compressibility. a.** $d\mu/dn$ measured as a function of filling factor ($\nu = n/(n_s/4)$) at temperatures $T = 4K, 50mK$ and $B_{||} = 0$, at the center of the sample ($\theta = 0.99°$, point in inset). A sharp asymmetric feature near $\nu = 2$ is indicated by an arrow. **b.** $d\mu/dn$ measured as a function of $\nu$ and spatial position $Y$ (along the same linecut as in fig. 1d, dashed line in inset) at $B_{||} = 12T$. The local twist angle changes along this linecut, as indicated on the right y-axis. The filling factor $\nu$ is determined by the local $n_s(Y)$ **c.** Selected $d\mu/dn$ traces from panel b with their corresponding $\theta$ lables, offset for clarity (purple curve is measured further along and outside the linecut in panel b and at $B_{||} = 10T$). **d.** The chemical potential $\mu$ as function of $\nu$, obtained by integrating the traces in panel c that correspond to $\theta = 0.99°$ and $1.05°$, offset for clarity. Near the CNP, in both curves $\mu$ increases in a $\sqrt{n}$-like fashion owing to the Dirac-like dispersion. In the larger $\theta$ curve this $\sqrt{n}$-like dependence reappears at all integer $\nu$'s, demonstrating the revival of Dirac-like behavior at the Fermi level (overall slope uncertainty is discussed in SI3). The horizontal black lines are equidistant in $\mu$. **e.** Zoom-ins on the $d\mu/dn$ jumps near various integer filling factors, showing a dip of $d\mu/dn$, just before the jump.

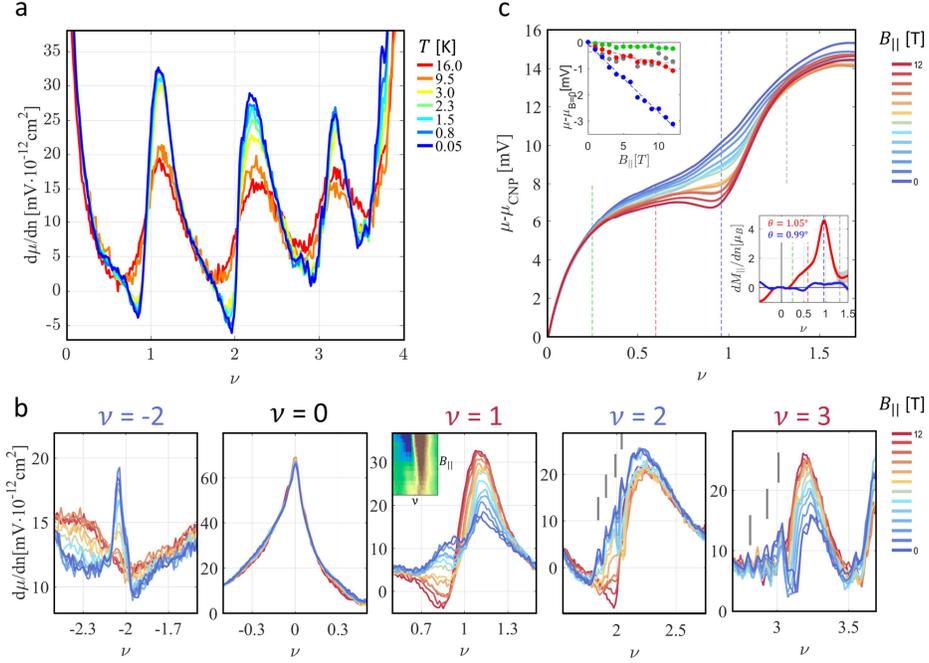

**Figure 3. Temperature and parallel magnetic field dependence near magic angle ($\theta = 1.05°$). a.** Temperature dependence: Measured $d\mu/dn$ as a function of $\nu$ at various temperatures, $T$, at $B_\parallel = 12T$. **b.** Parallel field dependence: $d\mu/dn$ measured as a function of filling factor around $\nu = -2, 0, 1, 2, 3$, at various values of $B_\parallel$ and $T$=50mK. The inset in the panel that corresponds to $\nu = 1$ replots the data in a colormap that emphasizes the position and magnitude of the $d\mu/dn$ dip (blue). **c.** $\mu$ as a function of $\nu$ for various values of $B_\parallel$, obtained from integrating the measured $d\mu/dn$. $\mu$ is referenced to the chemical potential at the CNP, $\mu_{CNP}$. Top inset: $\mu$ as a function of $B_\parallel$ at selected $\nu$'s (green, red, blue and gray dashed lines in main panel). Bottom inset: The differential magnetization $dM_\parallel/dn \equiv -d\mu/dB_\parallel$ extracted from a linear fit to the $B_\parallel$ dependence of $\mu$ at each $\nu$ measured at $\theta = 1.05°$ (red) and at $\theta = 0.99°$ (blue). While for $\theta = 0.99°$, $dM_\parallel/dn$ remains practically zero throughout, it peaks sharply near $\nu = 1$ for $\theta = 1.05°$, reaching a value of $4.5 \pm 0.3\mu_B$, signifying the buildup of spontaneous magnetization that starts around this peak. The gray region represents our error bar in determining $dM_\parallel/dn$, determined directly from the scatter of the data around the linear fit. This error grows linearly with $\nu$ due to the integration of parasitic capacitance and SET non-linearity errors (SI3).

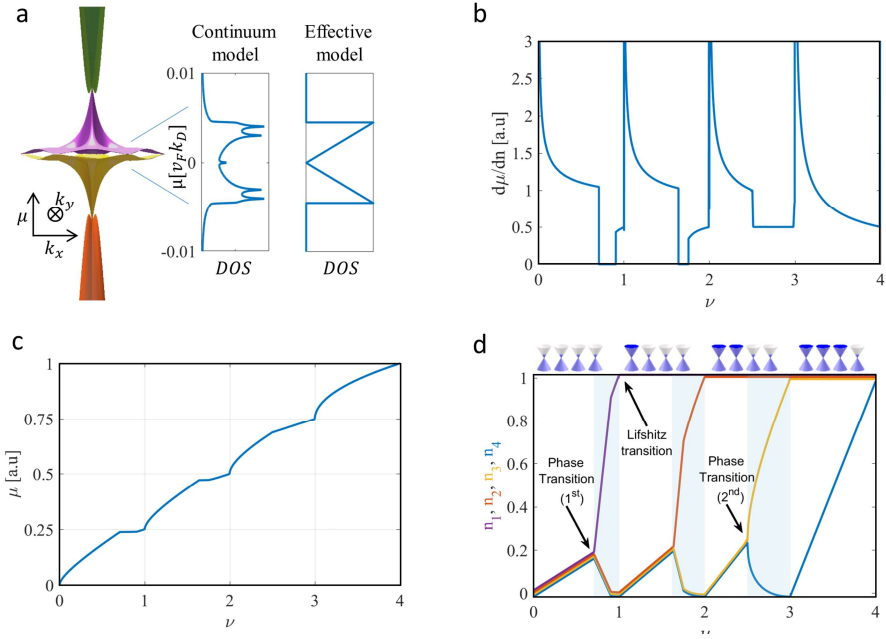

**Figure 4. Theory: Phase Transitions and Dirac Revivals**. **a.** Band dispersion (left) and density of states (DOS) (center) of the valence and conduction flat bands, calculated from a continuum model[7,8] for a twist angle of $\theta = 1.04°$. Our theory uses a simplified version of this DOS, with a triangular energy dependence (right), capturing the two essential features of the band structure: a) the gradual increase of the DOS from a Dirac-like dispersion at the CNP to a massive dispersion at the top and bottom band edges, and b) the consequential asymmetry of this DOS around the valence and conduction band centers. **b.** Hartree-Fock calculated $d\mu/dn$ vs. $\nu$ (see text and SI for details), showing apparent resemblance to the measured data (e.g. fig 2). **c.** The corresponding chemical potential, $\mu$, vs. $\nu$. **d.** The partial densities of individual flavors, $n_i$ ($i = 1..4$) as a function of $\nu$. The flavors start with four-fold degeneracy at charge neutrality, but then close to an integer $\nu$ experience a phase transition, after which one band takes all carrier from the other three (occurring in the shaded blue region). When the former band becomes full, the other three start refilling again from near the Dirac point. The beginning of this refilling corresponds to the Lifshitz transition (jump in $d\mu/dn$, panel b). This process repeats periodically near all integer filling factors, but each time with one less band, and can start with a 1$^{st}$ or 2$^{nd}$ order phase transitions (arrows). The first order transitions are followed by phase mixing regions where the partial densities interpolate linearly between those of the two phases. The flavor filling is illustrated on top.

**Acknowledgements:** We thank Uri Aviram, Allan H. Macdonald, Jonathan Ruhman, Hadar Steinberg, Senthil Todadri, Amir Yacoby and Eli Zeldov for their suggestions. Work at Weizmann was supported by the Leona M. and Harry B. Helmsley Charitable Trust grant, ISF grants (712539 & 13335/16), Deloro award, Sagol Weizmann-MIT Bridge program, the ERC-Cog (See-1D-Qmatter, no. 647413), the ISF Research Grants in Quantum Technologies and Science Program (994/19 & 2074/19), the DFG (CRC/Transregio 183), ERC-Cog (HQMAT, no. 817799), EU Horizon 2020 (LEGOTOP 788715) and the Binational Science Foundation (NSF/BMR-BSF grant 2018643). Work at MIT was supported by the National Science Foundation (DMR-1809802), the Center for Integrated Quantum Materials under NSF grant DMR-1231319, and the Gordon and Betty Moore Foundation's EPiQS Initiative through Grant GBMF4541 to P.J.-H. for device fabrication, transport measurements, and data analysis. This work was performed in part at the Harvard University Center for Nanoscale Systems (CNS), a member of the National Nanotechnology Coordinated Infrastructure Network (NNCI), which is supported by the National Science Foundation under NSF ECCS award no. 1541959. D.R-L acknowledges partial support from Fundaciò Bancaria "la Caixa" (LCF/BQ/AN15/10380011) and from the US Army Research Office grant no. W911NF-17-S-0001. K.W. and T.T. acknowledge support from the Elemental Strategy Initiative conducted by the MEXT, Japan, A3 Foresight by JSPS and the CREST (JPMJCR15F3), JST.

Data availability: The data that support the plots and other analysis in this work are available from the corresponding author upon request.

Contributions: U.Z., A.R., D.R-L., P.J-H. and SI designed the experiment. U.Z., A.R., performed the experiments. D.R-L. and Y.C. fabricated the twisted bilayer graphene devices. U.Z., A.R., and S.I. analyzed the data. R.Q., A.R., FvO, Y.O., A.S. and E.B. wrote the theory and performed the Hartree Fock calculations. K.W. and T.T. supplied the hBN crystals. U.Z., A.R., D.R-L., A.S., E.B., P. J-H. and S.I. wrote the manuscript, with input from all authors.

Supplementary materials for:

# Cascade of Phase Transitions and Dirac Revivals in Magic Angle Graphene

U. Zondiner[†], A. Rozen[†], D. Rodan-Legrain[†], Y. Cao, R. Queiroz, T. Taniguchi, K. Watanabe, Y. Oreg, F. von Oppen, Ady Stern, E. Berg, P. Jarillo-Herrero[*] and S. Ilani[*]

## Contents

## SI1. Twisted bilayer graphene device fabrication

The device was fabricated using a modified dry-transfer technique[1,2]. Monolayer graphene and hexagonal boron nitride (h-BN) were exfoliated on SiO$_2$/Si chips, and high-quality flakes were selected using optical microscopy and atomic force microscopy. The thickness of the top h-BN is 16nm and that of the bottom h-BN is 42nm, as determined by AFM measurements. We used a poly (bisphenol A carbonate) (PC)/polydimethylsiloxane (PDMS) stack on a glass slide mounted on a custom-made micro-positioning stage to pick up an h-BN flake at 110 °C, and then used the van der Waals force between h-BN and graphene to tear a graphene flake at room temperature. The separated graphene pieces were rotated manually by a twist angle of about 1.1°–1.2° and stacked together again, which resulted in a controlled twisted bilayer graphene (TBG) structure. The stack was then released on top of another h-BN flake, which was previously placed on top of a metallic gate. We did not perform any heat annealing after this step because we found that TBG tended to relax to Bernal- stacked bilayer graphene at high temperatures. The final device geometry was defined by using electron-beam lithography and reactive ion etching. Electrical connections were made to the TBG by Cr/Au edge-contacted leads[3].

Figure S1 shows an optical (panel a) and AFM (panel b) images of the device. The device has a large central region ($5\mu m \times 5\mu m$), with only two small bubbles visible in the AFM picture, allowing us to obtain consistent data over a large spatial range as well as observe systematic dependence of the physics on the twist angle, which changes gradually toward the edges of the device.

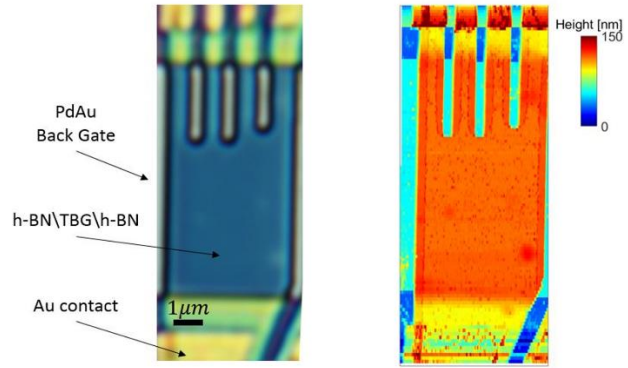

**Figure S1: Device Images. a.** Optical image of the device (scale bar 1um). The dark blue region corresponds to the h-BN/TBG/h-BN stack on top of a PdAu back gate which is visible on the right and left sides in dark green. Light green segments on the top and bottom correspond to parts of the stack that lie directly over the Si/SiO$_2$. The evaporated Cr/Au contacts are seen in yellow. **b.** AFM topographic image of the same area.

## SI2. Local compressibility measurement technique

We use a nanotube-based single electron transistor (SET)[4] to measure the local inverse electronic compressibility of the TBG, $d\mu/dn$. The measurement technique is identical to the one used previously[5–9]. The geometry of the measurement and its equivalent circuit are plotted in figures S2a,b. We park the SET at a fixed location above the TBG, apply an AC voltage excitation to the back gate, $\delta V_{BG} = 25mV$, at an acoustic tone $f_{BG}$ (in the range $80Hz - 1kHz$), keep the TBG grounded at this frequency, and measure the corresponding modulations of the source-drain current through the SET at this frequency, $\delta I_{SET}$. The back gate excitation modulates the density of the TBG via $\delta n_{TBG} = e^{-1}\delta V_{BG} C_{BG-TBG}$, where $C_{BG-TBG}$ is the capacitance between the back gate and the TBG, and $e$ is the electronic charge. Consequently, the chemical potential of the TBG also changes as $\delta \mu_{TBG} = (d\mu/dn)\delta n_{TBG}$. The TBG's electrochemical potential ($V_{TBG}$), electrostatic potential ($\phi_{TBG}$), and chemical potential ($\mu_{TBG}$) are tied together via the identity $V_{TBG} - \phi_{TBG} = \mu_{TBG}$. Since we keep the TBG grounded at the relevant frequency, $\delta V_{TBG} = 0$, then the change of the TBG's chemical potential translates directly to a change of its electrostatic potential: $\delta\phi_{TBG} = -\delta\mu_{TBG}$. This electrostatic potential gates the SET, modifying the average charge of its central 'island', which is formed in the suspended part of the nanotube between two p-n junctions. The modification to the SET charge is $\delta n_{SET} = e^{-1}\delta\phi_{TBG} C_{TBG-SET}$, where $C_{TBG-SET}$ is the capacitance between the area under study in the TBG and the SET's island (see fig. S2b). Owing to the Coulomb blockade phenomenon, a small change in $\delta n_{SET}$ causes a sharp change in the current through the SET, $\delta I_{SET}$, allowing us to detect small changes in the TBG's chemical potential. To experimentally normalize out the gain of the SET and its capacitance to the TBG, we apply simultaneously a second AC excitation on the contacts of the TBG, $\delta V_{TBG} = 1 - 5mV$, at a different frequency, $f_{TBG}$. At this frequency, the chemical potential of the TBG remains constant, $\delta\mu_{TBG} = 0$ (the gating due to the small excitation is negligible) so the electrochemical modulation that we impose translates directly to an electrostatic modulation, $\delta\phi_{TBG} = -\delta V_{TBG}$, which is similarly sensed by the SET. The ratio of the detected current modulations in response to these two excitations gives us directly:

$$\frac{\delta\mu_{TBG}}{\delta V_{BG}} = \left(\frac{\delta I_{SET}}{\delta V_{BG}}\right) / \left(\frac{\delta I_{SET}}{\delta V_{TBG}}\right)$$

The inverse compressibility of the TBG then follows from:

$$\left(\frac{d\mu}{dn}\right)_{TBG} = e\frac{\delta\mu_{TBG}}{\delta V_{BG}} / C_{BG-TBG}$$

The only parameter in this relation is the geometrical capacitance between the back gate and the TBG, which is known to $\pm 10\%$ in our experiment. It appears as a normalization factor in the relation above, and not as a large additive term as in capacitance measurements[10,11]. Thus, in contrast to capacitance measurements, where few percent error in the value of the geometrical capacitance can lead to hundreds of percent error in the extracted compressibility, here it translates only to a few percent error in the determined compressibility, making this measurement scheme quantitatively reliable.

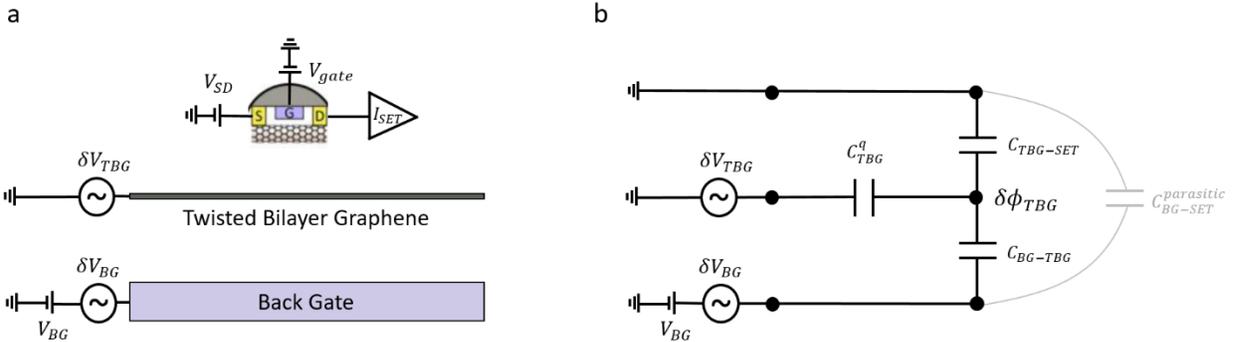

**Figure S2: Local compressibility measurement scheme**. **a.** Geometry of the measurement consisting of a back gate, twisted bilayer graphene (TBG), and a nanotube-based scanning single electron transistor (SET). A DC voltage is applied to the BG, $V_{BG}$, to tune the carrier density in the TBG. To measure the inverse compressibility of the TBG, two AC excitations at different acoustic frequencies are applied to the BG ($\delta V_{BG}$) and to the TBG ($\delta V_{TBG}$). The inverse compressibility is then obtained from the ratio of the measured SET current response at these corresponding frequencies (see text) **b.** The equivalent electrical circuit: $C_{BG-TBG}$ and $C_{TBG-SET}$ are the geometrical capacitances between the BG and TBG and the TBG and SET, and $C_{TBG}^q = dn/d\mu$ is the quantum capacitance of the TBG.

## SI3. Parasitic capacitance near sample edges and its mitigation

The analysis in the previous section was valid for measurements in the bulk of the sample. Near the TBG's edge, a direct line of sight is established between the SET and BG, leading to an additional parasitic capacitance (gray, figure S2b), adding an offset to the measured signal. This can clearly be seen in the measurement in figure S3a, which plots the penetration signal, $d\phi/dV_{BG}$, at a fixed carrier density ($V_{BG} = 0$), spatially imaged across the dashed black region demarcated in the right inset. Visibly, when the SET approaches an etched edge of the TBG there is an increased signal (red rim in the colormap), associated with the parasitic capacitance. Measurement along a line cut (figure S3b) shows that this parasitic background increases toward the etched edge over a spatial scale given by the scanning height of the SET above the TBG.

The parasitic capacitance adds a constant offset to inverse compressibility traces that are measured near sample edges. This is seen, for example, when we compare two $d\phi/dV_{BG}$ vs. $V_{BG}$ traces measured closer (red, fig S3c) and further away (blue, fig S3C) from the edge (colored dots in panel a show the measurement positions). By subtracting a constant offset from the first trace, we obtain a new trace (dashed black) that is practically identical to the one measured deeper into the sample's bulk, demonstrating that the parasitic capacitance adds merely a constant background. In the main text, for curves measured near the TBG's edge, we similarly subtract a constant background by demanding that the average of $d\mu/dV_{BG}$ over the entire $V_{BG}$ range of the measurement (typically from $\nu \approx -6$ to $\nu \approx 6$) is equal to the average measured in the bulk. This condition forces the chemical potential difference between the two $V_{BG}$ end-points of the measurement to remain fixed. By doing so, we still preserve the local information about the sawtooth features in $d\mu/dV_{BG}$ but lose the information about the bandwidth of the flat bands. Stated differently, our uncertainty in the overall background contribution in $d\mu/dV_{BG}$ generates an error in $\mu$ that grows linearly with $V_{BG}$, as represented by the gray region in figure S3d.

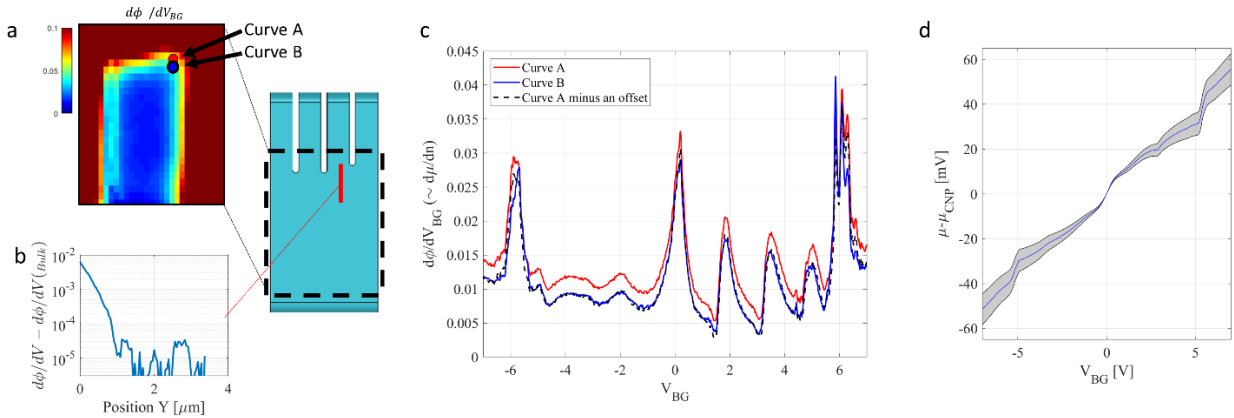

**Figure S3: Parasitic capacitance near sample edges and its mitigation**. **a.** Penetration signal, $d\phi/dV_{BG}$, spatially imaged along the area indicated by a black square in the right inset, at $V_{BG} = 0$. **b.** $d\phi/dV_{BG}$ measured as a function of the spatial coordinate, $Y$, along the red line in the right inset, where the average $d\phi/dV_{BG}$ in the bulk has been subtracted. **c.** Blue and red: traces of $d\phi/dV_{BG}$ vs. $V_{BG}$ measured at two different $Y$ coordinates (marked by dots in panel a). Dashed black curve plots the red curve after subtraction of a constant offset. **d.** $\mu$ vs. $V_{BG}$ trace (blue) obtained from integration of measured $d\phi/dV_{BG}$. Gray region shows the error bar in this integration due to parasitic capacitance uncertainty of $(d\phi/dV_{BG})_{parasitic} = 10^{-3}$.

## SI4. Determining the map of local twist angles from compressibility measurements

There are two independent types of disorder in TBG: charge disorder, and twist angle disorder. Charge disorder shifts both the back gate voltage that corresponds to the charge neutrality point (CNP), $V_{CNP}$, and the voltages that correspond to full valence and conduction flat bands, $V_{\pm n_s}$, by the same amount. Angle disorder, on the other hand, does not shift $V_{CNP}$ but changes $n_s$, and correspondingly shifts $V_{-n_s}$ and $V_{+n_s}$ in opposite directions. These voltages thus allow us to independently determine the twist angle and charge offset at any spatial point.

The charge disorder in our sample, extracted by tracking $V_{CNP}$ in space, is $\sim 5 \cdot 10^{10} cm^{-2}$, whereas the measured shifts in $n_s$ are $\sim 7 \cdot 10^{11} cm^{-2}$. This shows that the dominant disorder is angle disorder and implies that to determine $\theta$ we can neglect $V_{CNP}$ shifts and merely follow $V_{+n_s}$.

Figure S4 shows $d\mu/dn$ measured as a function of the $X$ spatial coordinate and $V_{BG}$, at few $Y$ coordinates (complementary to Panel 1d in the main paper). At the center of the sample, we observe that $\theta$ is practically constant ($V_{+n_s} = 5.4V$). Going toward the top edge of the sample, we see that $V_{+n_s}$ shifts up in steps. The map of $\theta(X,Y)$ in fig. 1e in the main text is obtained by following $V_{+n_s}$ in a similar, but denser, three dimensional dataset.

It is important to note that since our SET is scanned at a finite height above the sample ($h \sim 500nm$), which can be comparable in magnitude to the size of the domains, it often happens that when parked at a single spatial position, the SET measures the $d\mu/dn$ signal also from adjacent domains. This is shown, for example, in figure S4b; its top panel replots fig. 1d from the main text, showing $d\mu/dn$ measured as a function of $V_{BG}$ and $Y$. In this panel, we marked the observed $V^i_{+n_s}$ values. The bottom panel plots the magnitude of $d\mu/dV_{BG}$ at these specific $V^i_{+n_s}$ back gate voltages, as a function of $Y$. Visibly, when the signal from one domain increases, the signals from its neighboring domains decrease. However, at a given point in space, in addition to the $d\mu/dn$ emanating from the domain directly beneath the SET we also see tails of $d\mu/dn$ coming from adjacent domains. Thus, at a given position, the SET can often measure multiple $V_{+n_s}$ peaks.

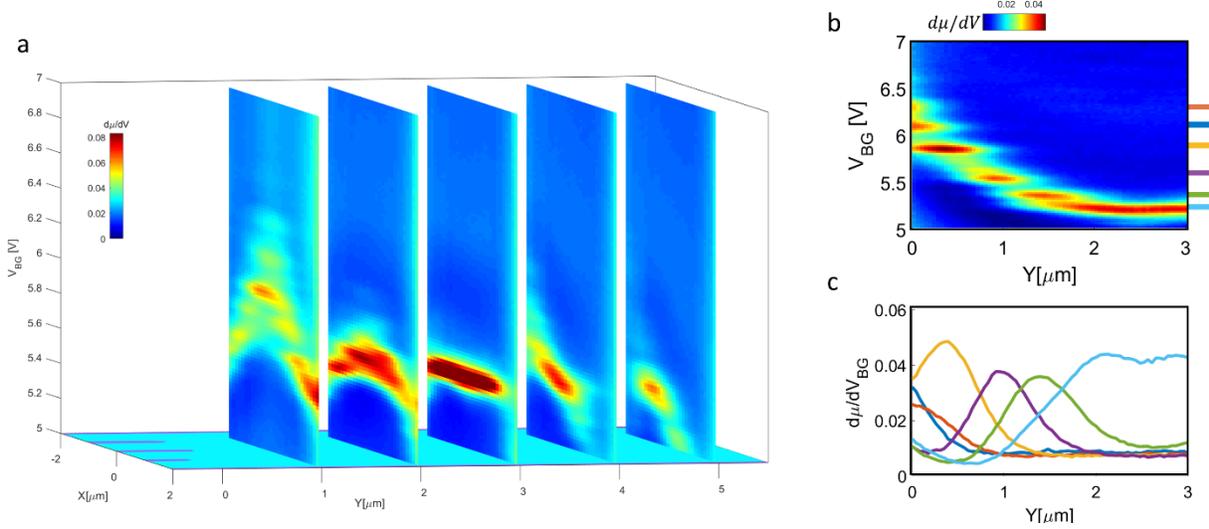

**Figure S4: Determining the local twist angle from compressibility measurements a.** $d\mu/dV_{BG}$ measured as a function of $V_{BG}$ and spatial coordinate $X$, at several $Y$ positions, superimposed over the schematics of the device. **b.** Replot of fig. 1d from the main text, showing $d\mu/dV_{BG}$ measured as a function of $V_{BG}$ and $Y$. Colored lines on the right mark a discrete set of back gate voltages, $V_{BG} = V^i_{+n_s}$, at which we observe a peak in $d\mu/dV_{BG}$ that corresponds to a full flat band. **c.** Plots of the $d\mu/dV_{BG}$ value at the back gate voltages $V_{BG} = V^i_{+n_s}$ (color of the trace corresponds to the color labeling in panel b). Visibly, the various inverse compressibility peaks, which correspond to domains with different twist angle, reach a maximum at different $Y$ locations. We can also see that due to the finite spatial resolution of our measurement, given primarily by the scanning height of the SET above the TBG, at a given spatial point, in addition to the signal from the domain underneath the SET, we measure also inverse compressibility tails from neighboring domains.

### SI5. Two-probe transport between various contact pairs

Measured transport between various contact pairs provides a rough estimate of the spatial inhomogeneity of the twist angle in the sample, which can then be compared with the more detailed local information provided by the scanning SET. Fig. 1b in the main text showed a 4-probe transport measurement performed with the four top edge contacts (labeled C1-C4 in fig. S5a). This measurement exhibited resistance increase at half-filled conduction and valence flat bands ($\nu = \pm 2$) as well as superconducting-like resistance drops at the electron doped side of $\nu = 2$ and the hole doped side of $\nu = -2$. These observations are consistent with the phenomenology seen previously in transport measurements of samples with a twist angle near the MA [12,13], and implies the existence of a region with MA phenomenology at the top edge of our sample between contacts 2 and 3. However, the finite resistance observed both in the correlated insulating peaks, and the superconducting-like resistance drops suggest, that this MA region is in series/parallel to other non-MA regions.

This picture is validated by two probe measurements done between various contact pairs (figure S5). In these measurements one probe is always the contact C5 and the other is chosen from the contacts C1-C4 (panels a-d). In all cases, we observe resistive peaks at full filling, although the corresponding gate voltage varies between the measurements, reflecting a twist angle change of $\Delta\theta/\theta \sim 10\%$. Two traces (panels a and d) show a single full band peak, whereas the other two (panels b and c) exhibit multiple full-band peaks, indicating that the transport path in the latter crosses multiple domains with different $\theta$'s. The measurements corresponding to contacts C1,C3,C4 barely show any resistive features at $\nu = \pm 2$,

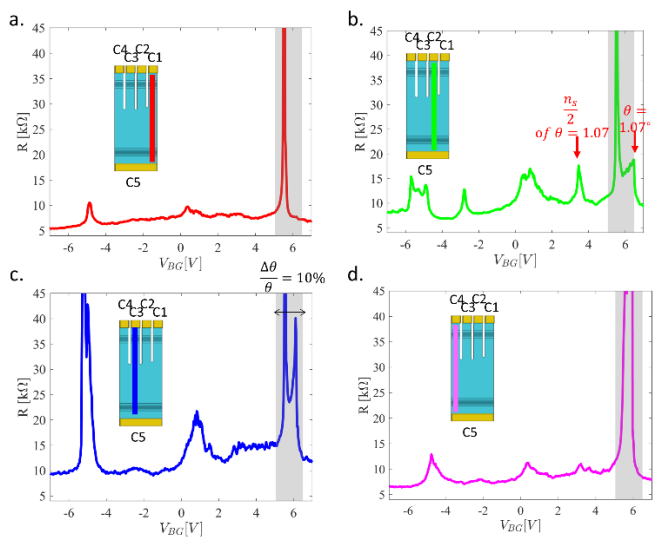

**Figure S5: Two-probe resistance measurements across various contact pairs.** Panels **a-d** show the two probe resistance between contact C5 and contacts C1 to C4 (as illustrated in insets), measured as a function of $V_{BG}$. In all cases we observe resistive peaks that correspond to full flat bands. Their corresponding density, $n_s$, varies between the various measurements, reflecting a 10% spatial variation in the twist angle. In panel b a full band peak that corresponds to $\theta = 1.07°$ appears, and concomitant with it clear insulating peaks at the corresponding $\pm n_s/2$ also appear.

consistent with the fact that the angles obtained from their full band peak are lower than the MA.

In contrast, transport to contact C2 shows nicely resolved peaks at $\nu = \pm 2$. These peaks appear in correlation with the emergence of a full band peak at $V_{BG} = 6.2V$, which corresponds to $\theta = 1.07°$. Note that a full band peak that corresponds to lower $\theta$'s still dominates this trace, demonstrating that lower $\theta$ regions still contribute significantly in this conduction path. Taken together, these two-probe measurements demonstrate that the region with near-MA twist angle is located above contact C2, which is also roughly the region probed in the four probe measurements in fig. 1b. Moreover, these results nicely match the local compressibility measurements. For example, the multiple full band peaks in the transport between C2 and C5 are readily explained by the angle maps generated by the SET (fig. 1e).

## SI6. Compressibility measurements of an independent spatial region with a near-magic twist angle.

To demonstrate that the Dirac revivals physics described in the main text is robust and reproducible, we utilize our scanning ability to see whether the same physics appears in spatially independent regions in the sample that have near MA twist angle.

Figure S6 shows the twist angle map across our sample. The near MA region that was explored in the main text appears at the top of this scanning window, together with a characteristic $d\mu/dn$ vs. $\nu$ trace (panel a), measured at the position of the black star and with $B_{\parallel} = 12T$. This curve shows the prototypical sawtooth-like features at $\nu = 1,2,3$ as described in the main text. In contrast, when we measure $d\mu/dn$ vs. $\nu$ at the center of the sample (blue star), where $\theta = 0.99°$, we see (panel b) that the asymmetric sawtooth-like feature at $\nu = 1$ completely disappears, being replaced by a symmetric hump. In fact, throughout the bulk of the sample (e.g. along the dashed blue line), we observe the absence of an asymmetric $\nu = 1$ feature, in correlation with a low twist angle. However, near the bottom of the sample there is another region where the angle climbs back toward the MA. Measurement of $d\mu/dn$ vs. $\nu$ in this region (panel c, measured at the location of the purple star) shows a clear re-emergence of a highly pronounced sawtooth-like feature at $\nu = 1$, which is fully consistent with the measurement in the top part of the sample. The re-emergence of this peak happens concomitantly with the appearance of a full band peak that corresponds to a twist angle of $\theta = 1.04°$. This data demonstrates the reproducibility of the sawtooth features and shows that they appear whenever $\theta$ approaches the MA.

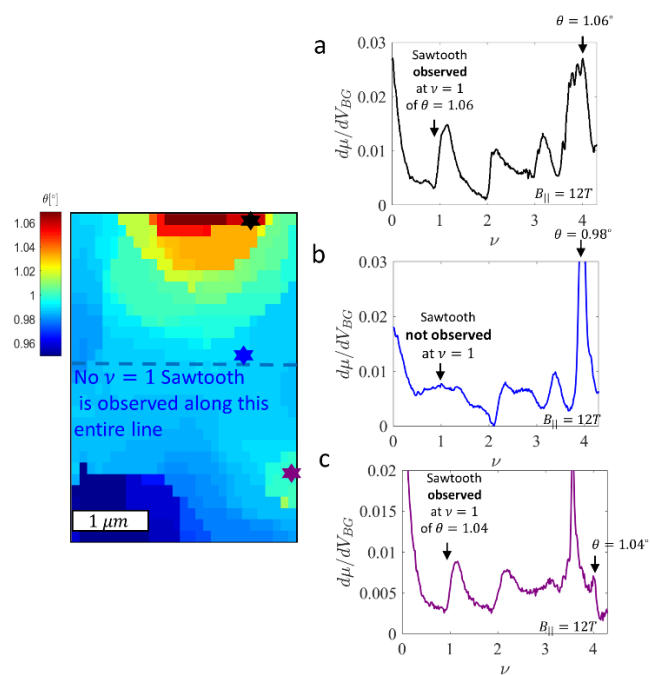

**Figure S6: Observing similar Dirac revival sawtooth features in two spatially separated near-MA regions.** Left: map of the twist angles in the sample (taken from fig. 1e). **a-c** Traces of $d\mu/dn$ vs. $\nu$ measured at three points in the sample at $B_{\parallel} = 12T$. The central curve is characteristic of traces measured in the bulk of the sample (e.g. throughout the entire dashed blue line). It corresponds to a twist angle of $\theta = 0.99°$, and does not exhibit an asymmetric sawtooth feature at $\nu = 1$. The top and bottom curves correspond to two separate regions in the sample in which a full band peak corresponding to a twist angle neat the MA appears. Concomitantly, a strong asymmetric sawtooth feature at $\nu = 1$ also appears, demonstrating that this physics is robust and driven primarily by the twist angle.

## SI7. Upper bound on the size of a possible thermodynamic energy gap at half filling from the compressibility data

Transport measurements[13–15] of MATBG have observed resistive behavior near integer fillings that were associated with the formation of correlated insulator state. Temperature dependence measurements displayed activated behavior which corresponds to small energy gaps, $0.14 mV - 0.9 mV$. STM measurements[16,17], on the other hand, have observed an order-of-magnitude larger gaps at $\nu$=2, reaching a value of $8 mV$. Since STM measures local information, it does not suffer from disorder smearing as does transport in macroscopic samples, and therefore can better measure the local gaps. On the other hand, tunneling experiments can also observe a completely different type of gap, a Coulomb gap, which originates from the suppression of the tunneling process itself and is unrelated to the presence of a correlated gap in the system.

Scanning compressibility measurements provide a suitable tool to mitigate the above limitations. Firstly, similar to STM these measurements are local and thus can isolate the effects of charge and angle disorder. Second, the electronic compressibility is a thermodynamic property that probes the energetics of the system directly. Thus, in contrast to tunneling experiments, it does not exhibit Coulomb gaps, but should show the presence of a thermodynamic gap directly. Finally, in contrast to activation measurements where the temperature needs to be changed over a significant range to determine the size of the gap, here the gap can be measured directly at the base temperature. This is especially useful for small gaps that leave a rather limited measurable temperature range for activation and also for gaps whose magnitude changes with temperature.

A gap has a straightforward thermodynamic signature in compressibility measurements. Since a gap involves a jump in the chemical potential, $\mu$, the derivative of this potential, $d\mu/dn$, should exhibit a peak. Indeed, we observe exactly such peaks at full band fillings, as shown for example in figure S7, which plots the inverse compressibility measured at the sample center. The area under the $d\mu/dV_{BG}$ vs. $V_{BG}$ peaks (green in the figure) gives directly and without any free parameter the size of the gap, which for the case of the trace in the figure equals $15 meV$. The width of the peak is determined by the disorder.

In contrast to the behavior at full flat band, near $\nu = 2$ the inverse compressibility does not show any indication of a peak but instead exhibits an asymmetric sawtooth-like feature. It is still possible, however, to have a sufficiently small thermodynamic gap at $\nu = 2$ whose corresponding inverse compressibility peak is unresolvable within our measurement. To obtain an upper bound on such a possible gap, we overlay near $\nu = 2$ in fig. S7 the peaks that would correspond to gaps of $E_g = 0.3, 1, 2 mV$. These peaks are rescaled duplicates of the peak measured at full filling, where the density axis has been divided by a factor of 2 to map the physics at full filling onto half filling and the height of this peak has been rescaled down by $2E_g/15mV$, to make the area under the peak equal $E_g$. This procedure gives the most realistic representation of the expected peaks because, by construction, it gives the smearing obtained by the actual disorder measured at the same spatial point (originating primarily from angle disorder). Comparing the peaks with the measured data yields an upper bound of $\sim 0.5 mV$ for a possible thermodynamic gap. This value is consistent with the gaps measured by activation in transport experiments, but is significantly smaller than the gaps observed by STM.

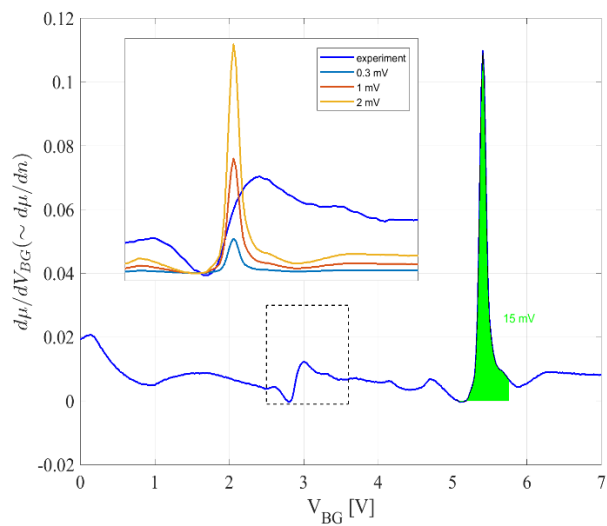

**Figure S7: Upper bound on the size of a possible thermodynamic gap.** Main panel: $d\mu/dV_{BG}$ vs. $V_{BG}$ measured at the center of the sample. The peak at $V_{BG} = 5.4V$ corresponds to the band gap at $n_s$, whose magnitude, obtained by integrating this curve, is $15 mV$. Inset: zoom-in on the sawtooth-like feature in compressibility near $\nu = 2$. The overlayed yellow, red, and light blue curves are compressed and rescaled duplicates of the peak at full filling, to reflect possible gaps whose magnitude is $E_g = 2mV, 1mV, 0.3mV$ (see text). The comparison gives an upper bound of $\sim 0.5 mV$ on a possible thermodynamic gap.

## SI8. Parallel magnetic field dependence of compressibility for $\theta = 0.99°$

In fig. 3b of the main text, we presented the $B_\parallel$ dependence of $d\mu/dn$ vs. $\nu$ curves taken at a near-magic twist angle ($\theta = 1.05°$). Here we present additional $B_\parallel$ dependence data at the center of the sample, where $\theta = 0.99°$. Figure S8 presents the evolution of the $d\mu/dn$ traces for the range $B_\parallel = 0T$ to $12T$. As shown in the main text, for $\theta$ below the MA sawtooth features are observed only near $\nu = 2,3$. Moreover, we observed that at lower $\theta$ the $\nu = 3$ sawtooth is not 'fully developed' in the sense that it actually does not start precisely at $\nu = 3$ but closer to $\nu = 3.5$. Similarly to the near-MA data in the main paper, also here we see that $B_\parallel$ operates oppositely on the sawtooth features at even and odd fillings. The $\nu = 3$ sawtooth increases in magnitude and gets closer to integer filling with increasing $B_\parallel$, whereas it slightly decreases in magnitude and gets further away from integer filling for $\nu = 2$. Moreover, while the measurement in fig. 3b near $\nu = 2$ and $\nu = 3$ had shadow sawtooth features, originating from neighboring smaller-angle domains that are also detected by the SET, here the $\theta$ has no spatial variation so the SET measures only a single $\theta = 0.99°$ domain. Correspondingly the sawtooth features at $\nu = 2$ and $\nu = 3$ do not have any shadow features, allowing to observe the magnetic field dependence very cleanly.

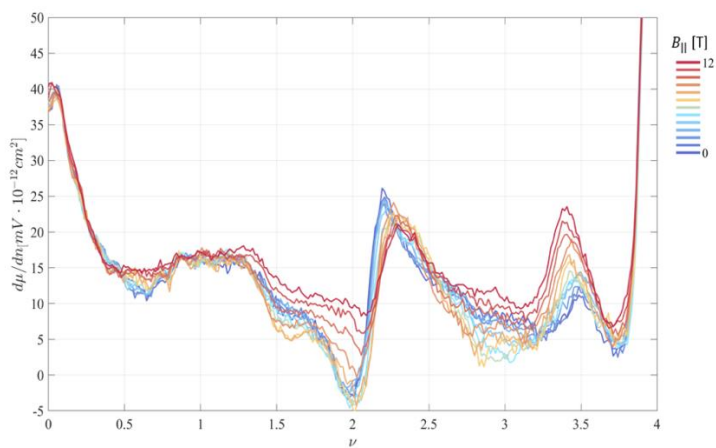

**Figure S8: Parallel magnetic field dependence at $\theta = 0.99°$.** $d\mu/dn$ measured as a function of $\nu$, for various values of $B_\parallel$, at T=50mK.

## SI9. Temperature scale of the observed Dirac revivals

In fig. 3a, we presented the measured temperature dependence of the sawtooth features in $d\mu/dn$, showing that up to the highest temperature achievable in our scanning microscope setup ($T = 16K$), the phenomenon remains significant. We can obtain a rough estimate of the temperature scale of the phenomenon by plotting the amplitude of the steps in $d\mu/dn$ around the $\nu = 1,2,3$ sawtooth features. The amplitude is determined by $A_{sawtooth}^{\nu=1,2,3} = \max\limits_{\nu\in[i-0.4,i+0.3]}(d\mu/dn) - \min\limits_{\nu\in[i-0.4,i+0.3]}(d\mu/dn)$ and plotted as a function of $T$ in fig. S9. Linear fits to the measured points show that the phenomenon is expected to disappear roughly around $T \approx 30K$.

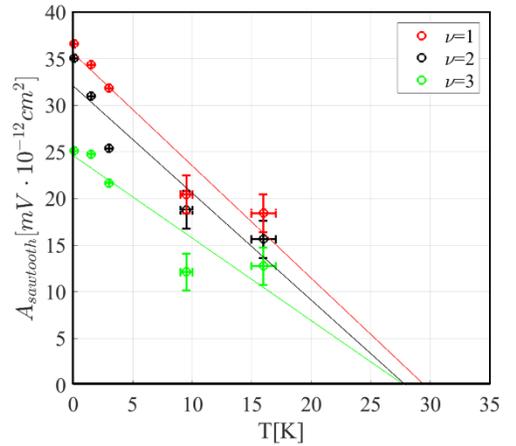

**Figure S9: Temperature scale of the observed sawtooth features in $d\mu/dn$.** The amplitude of the sawtooth features around $\nu = 1,2,3$ is extracted from fig. 3a in the main text as the difference in $d\mu/dn$ before and after the sawtooth feature, namely, $A_{sawtooth}^{\nu=1,2,3} = \max\limits_{\nu\in[i-0.4,i+0.3]}(d\mu/dn) - \min\limits_{\nu\in[i-0.4,i+0.3]}(d\mu/dn)$. This is plotted as dots for the various measurement temperatures. The lines show the linear fit to the amplitudes at $\nu = 1,2,3$ and extrapolate to zero amplitude around $T \approx 30K$.

## SI10. Theory: Model and mean-field theory

Here, we describe a simple model used to describe interaction effects in the nearly-flat bands of twisted bilayer graphene (TBG). The single-particle states are described by the Bistritzer-MacDonald model[18]. The Hamiltonian is given by:

$$H = \sum_{\mathbf{k},\alpha,m} (\varepsilon_{\alpha m \mathbf{k}} - \mu)\psi^\dagger_{\alpha m \mathbf{k}}\psi_{\alpha m \mathbf{k}}$$
$$+ \frac{1}{2}\sum_{\alpha,\beta} \int d^2r d^2r' V(\mathbf{r}-\mathbf{r}')\psi^\dagger_\alpha(\mathbf{r})\psi^\dagger_\beta(\mathbf{r}')\psi_\beta(\mathbf{r}')\psi_\alpha(\mathbf{r}) - E_0 + \mu N_0 \quad (1)$$

Here, $\alpha$ represents the spin/valley index such that the states $\{K\uparrow, K\downarrow, K'\uparrow, K'\downarrow\}$ are labelled as $\alpha = \{1,2,3,4\}$, respectively. The operator $\psi^\dagger_{\alpha m \mathbf{k}}$ creates an electron in a Bloch state in band $m$, with flavor index $\alpha$ and quasi-momentum $\mathbf{k}$, whose band dispersion is $\varepsilon_{\alpha m \mathbf{k}}$. $\psi_{\alpha m \mathbf{k}}$ is related to the field operators in real space by $\psi_\alpha^\dagger(\mathbf{r}) = \sum_{m,\mathbf{k}} \phi_{\alpha m \mathbf{k}}^*(\mathbf{r}) \psi_{\alpha m \mathbf{k}}^\dagger$, where $\phi_{\alpha m \mathbf{k}}(\mathbf{r})$ is the Bloch wavefunction. Henceforth, we limit the summation over $m$ to the valence and conduction nearly flat bands, assumed to be separated from the rest of the spectrum by a gap. Note that $\varepsilon_{\alpha m \mathbf{k}}$ depends on the valley index, but not on the spin. $V(\mathbf{r} - \mathbf{r}')$ is the effective interaction between electrons. For later convenience, we have subtracted $E_0$, defined as the ground state energy at the charge neutrality point, and added $\mu N_0$, where $N_0$ is the number of electrons corresponding to overall charge neutrality (four electrons per moiré unit cell).

The Hamiltonian (1) has a $U(2) \times U(2)$ symmetry, associated with the conservation of spin and charge in each valley[19]. We neglect lattice-scale exchange interactions that break this symmetry into $U(1) \times U(1) \times SU(2)$ (charge conservation within each valley and total spin conservation). These terms are expected to be small in proportion to the ratio between the atomic lattice spacing to the moiré unit cell size[19], similar to the situation in the quantum Hall regime in graphene[20].

The Hamiltonian (1) is, of course, very difficult to solve in the physically relevant regime where the interaction strength is comparable to the bandwidth. Impressive progress has nevertheless been achieved in the strong coupling limit[21]. In particular, the possibility of ferromagnetic states with spontaneously broken spin/valley symmetry has been emphasized[22]. Based on these insights, and guided by our experimental results, we focus on the effects of the inter-flavor exchange interactions, and make the following approximations:

1. We assume that the interaction is short ranged in real space, and hence, due to the Pauli principle, electrons with the same valley and spin indices do not interact. . The effect of the long-ranged part of the interaction is discussed in a separate subsection below.
2. Writing the interactions in terms of the Bloch states, we neglect the dependence of the interaction matrix element on the spin, valley, and band indices. I.e., we approximate the interaction term as $\frac{U}{2N} \sum_{\alpha \neq \beta} \sum_{\{m_i\},\{\mathbf{k}_i\},\mathbf{G}} \delta_{\mathbf{k}_1+\mathbf{k}_2-\mathbf{k}_3-\mathbf{k}_4+\mathbf{G}} \psi_{\alpha m_1 \mathbf{k}_1}^\dagger \psi_{\beta m_2 \mathbf{k}_2}^\dagger \psi_{\beta m_3 \mathbf{k}_3} \psi_{\alpha m_4 \mathbf{k}_4}$, where $N$ is the number of moiré unit cells in the system and $\mathbf{G}$ is a reciprocal moiré lattice vector, with a single interaction parameter $U$.
3. Finally, we perform a Hartree-Fock analysis on the resulting Hamiltonian, in which we allow the $U(2) \times U(2)$ valley/spin symmetry to be broken spontaneously by populating the different valleys and spins differently.

These approximations clearly cannot describe some important physical effects, such as the renormalization of the bands due to the interactions, which has been argued for theoretically[23–25] and seems to be significant experimentally[16,17,26], and the possibility of spontaneous breaking of spatial symmetries, such as the threefold rotational symmetry[16,27–29]. Nevertheless, as we elaborate below, this simple description captures many of the salient features seen in our measurements of the compressibility, as well as key features seen in other experiments.

With these simplifying assumptions, we use a trial wavefunction $|\Psi_{MF}\rangle$ which is the ground state of the variational Hamiltonian

$$H_{MF} = \sum_{\mathbf{k},\alpha,m} (\varepsilon_{\alpha m \mathbf{k}} - \mu_\alpha) \psi_{\alpha m \mathbf{k}}^\dagger \psi_{\alpha m \mathbf{k}}, \tag{2}$$

and minimize $E_{MF}(\{\mu_\alpha\}) \equiv \langle \Psi_{MF}|H|\Psi_{MF}\rangle$. The mean-field grand potential[1] per moiré unit cell can be written as

$$\frac{\Phi_{MF}(\{\mu_\alpha\})}{N} = \sum_\alpha E_k(\mu_\alpha) + \frac{U}{2}\sum_{\alpha\neq\beta} \nu(\mu_\alpha)\nu(\mu_\beta) - \mu\sum_\alpha \nu(\mu_\alpha), \quad (3)$$

where $N$ is the number of moiré unit cells, while

$$\nu(\varepsilon) = \int_0^\varepsilon d\varepsilon' \rho(\varepsilon') \quad \text{and} \quad E_k(\varepsilon) = \int_0^\varepsilon d\varepsilon' \, \varepsilon' \rho(\varepsilon') \quad (4)$$

For convenience, we measure both the density and the kinetic energy relative to the charge neutrality point, $\mu = 0$. We have introduced the density of states (DOS):

$$\rho(\varepsilon) = \frac{1}{N}\sum_{\mathbf{k},m} \delta(\varepsilon - \varepsilon_{\alpha m \mathbf{k}}). \quad (5)$$

Note that $\rho(\varepsilon)$ does not depend on the flavor index, since although the two valleys have different dispersion relations, their densities of states are the same. Fixing $\mu$ and minimizing the energy with respect to $\{\mu_\alpha\}$, we obtain the total filling per unit cell, $\nu(\mu) = \sum_\alpha \nu(\mu_\alpha)$. The inverse compressibility, which is the quantity measured in our experiment, can then be obtained: $\kappa^{-1} = d\mu/dn$, where $n$ is the density of electrons per unit area, $n = \nu/a^2$ with $a^2$ being the area of the moiré unit cell.

It is convenient to invert Eqs. (4) and express $E_k$ as a function of $n$ rather than $\varepsilon$. The variational energy (3) is then written as a function of $\nu_\alpha \equiv \nu(\mu_\alpha)$. Minimizing with respect to $\nu_\alpha$, we obtain

$$\frac{\partial E_k}{\partial \nu_\alpha} + U\sum_{\beta\neq\alpha} \nu_\beta - \mu = 0. \quad (6)$$

Solving these four coupled equations gives $\nu_\alpha(\mu)$. We can then differentiate Eq. (6) with respect to $\mu$ and sum over all flavors, obtaining the compressibility:

$$\frac{dn}{d\mu} = \frac{\bar{\rho}}{1+\bar{\rho}U}\frac{1}{a^2}, \quad (7)$$

where $\bar{\rho} = \sum_\alpha \frac{\rho_\alpha}{1-\rho_\alpha U}$ and $1/\rho_\alpha \equiv d^2 E_k/d\nu_\alpha^2$ is the inverse single-particle DOS of flavor $\alpha$. When $\rho_\alpha U = 1$, $\bar{\rho}$ diverges, which corresponds to the Stoner criterion.

## SI11. Theory: Results for Linear Density of States

With our assumptions, the only input needed from the single-particle band structure is the density of states $\rho(\varepsilon)$. Theoretically, it is expected that near charge neutrality, the bands have a Dirac spectrum; the Dirac points are stable as long as the symmetry under $C_2\mathcal{T}$ (time reversal followed by a rotation by $\pi$ around an axis perpendicular to the TBG) is maintained. Thus, we assume that $\rho(\varepsilon) \propto |\varepsilon|$ near $\varepsilon = 0$. We begin our analysis with the following simple model for the DOS:

$$\rho(\varepsilon) = \frac{2|\varepsilon|}{W^2}\Theta(W - |\varepsilon|). \quad (8)$$

---
[1] This is the grand potential rather than the energy, since we are working at a fixed chemical potential.

Here, $2W$ is the combined bandwidth of the two narrow bands. At the edges of the flat bands, $\varepsilon = \pm W$, we assume that the DOS drops discontinuously to zero, corresponding to a quadratic dispersion near the band top and bottom. Later, we shall see that a more complicated model for $\rho(\varepsilon)$, that includes the van Hove singularities expected near the middle of the conduction and valence bands, does not change the results qualitatively in the relevant regime of intermediate interaction strengths. On the other hand, we find that it is imperative that a large density of states exists close to the top of the conduction band and the bottom of the valence band.

With (8), we can express $E_k$ in terms of $\nu$ rather than $\mu$:

$$E_k(\nu) = \frac{2}{3}W|\nu|^{3/2} \quad (-1 \leq \nu \leq 1). \tag{9}$$

such that the self-consistent equation (6) can be expressed as set of quadratic equations

$$W\,sign(\nu_\alpha)\sqrt{|\nu_\alpha|} + U(\nu - \nu_\alpha) = \mu, \text{ with } \nu = \sum_\alpha \nu_\alpha \text{ and } -1 \leq \nu_\alpha \leq 1. \tag{10}$$

The phase diagram of the model as a function of $\mu$ and $U$, obtained by minimizing the mean-field energy (3) with the kinetic energy given by (9), is shown in figure S10a. Figure S10b shows the inverse compressibility $d\mu/dn$ as a function of filling $\nu$ and $U$. As we will discuss below, it is rather rich and includes first and second order phase transitions as a function of the parameters $\mu$ and $U$. The mean-field equations can mostly be solved analytically, as we describe below.

For $U = 0$ the bands are filled for the four flavors in a symmetric fashion. For $\mu < -W$, the bands are empty ($\nu_\alpha = -1$ for all $\alpha$). As $\mu$ increases, the neutrality point is reached at $\mu = 0$ ($\nu_\alpha = 0$ for all $\alpha$), and the bands become completely filled at $\mu = W$ ($\nu_\alpha = 1$ for all $\alpha$). Using the minimization condition (6) the solutions are given, in terms of the total density $4n_\alpha = n$ and $\mu$, by:

|  | $\nu_\alpha$ | $dn_\alpha/d\mu$ | $\kappa^{-1}/a^2 = (d\mu/dn)/a^2$ |
|---|---|---|---|
| $W < \mu$ | 1 | 0 | $\infty$ |
| $-W < \mu < W$ | $(\mu/W)^3/|\mu/W|$ | $2|\mu|/W^2 = \sqrt{|\nu|}/W$ | $W^2/(8|\mu|) = W/(4\sqrt{|\nu|})$ |
| $\mu < -W$ | $-1$ | 0 | $\infty$ |

The inverse compressiblity, $\kappa^{-1}$, jumps to infinity when the bands are fully occupied or completely empty, and diverges at the charge neutrality point, where $\nu = 0$ and $\mu = 0$.

We expect that a similar symmetric solution still holds for small $U$. The occupation of each band is given by the solution of the quadratic equation (10). Substituting $\nu_\alpha = \pm 1$ into this condition, we find that the interaction $U$ stretches the region in the $\mu - U$ plane in which the bands are partially filled to $-W - 3U < \mu < W + 3U$. The boundary between this region and the region of completely filled bands is clearly seen in figure S10.

Figure S10 shows that below a critical value of $U$, $U_{c,0} = W/2$, the ground state remains fully symmetric at all densities. For $U > U_{c,0}$, symmetry broken phases start appearing. This can be understood by examining the expression for $\bar{\rho}$ after Eq. (7). It indicates that an instability will occur at $\rho_\alpha U = 1$. According to Eq. (8), the maximal value of $\rho_\alpha$ is $2/W$. Thus, the symmetric state is stable for $U < W/2$. For $U = W/2$ the bands become completely filled at $\mu = W + 3U = 2.5W$, so that the symmetry broken phases emanate from the point ($\mu = 2.5W, U = 0.5W$) in the phase diagram. We find numerically that the symmetry broken phases extend to lower densities as $U$ increases. In the symmetry broken phase with the lowest density, the flat bands are fully occupied for one flavor, while they are partially and equally occupied for the other three flavors. The shape of the corresponding phase boundary can be obtained by equating the energies of the two competing phases, and will be derived below.

For intermediate values of $U$, comparable with the bandwidth, we expect that the flavors will occupy the bands sequentially. As one flavor gets filled, it pushes the band energies for the other flavors towards higher energies, completely breaking the symmetry between the four flavors. When $U$ is large enough, such a symmetry broken phase will be present at the charge neutrality point $\mu = 0$. In this case, two flavors are completely filled (e.g., $\nu_3, \nu_4 = 1$), while the other two remain completely empty ($\nu_1, \nu_2 = -1$). Comparing the energies of the symmetric and symmetry broken configurations, $E_{MF}(\{-1,-1,1,1\}) = 8W/3 - 2U$ and $E_{MF}(\{0,0,0,0\}) = 0$, we find that the ground state remains the symmetric state $\nu_\alpha = 0$ for $U < U_{c,1} = 4W/3$, whereas it is symmetry broken with $\{\nu_\alpha\} = \{-1,-1,1,1\}$ (or any permutation thereof) for $U > U_{c,1}$. At $U = U_{c,1}$ there is a first-order transition between these two states. Since the experimental results are not compatible with a fully polarized state at charge neutrality (which would be an incompressible state with a substantial gap), we focus on the regime $U < U_{c,1}$.

Let us focus on the intermediate interaction strength regime, $U_{c,0} < U < U_{c,1}$, where the bands are filled sequentially. Figure S11 shows a cut through the phase diagram with $U = 1.2W$. As $\mu$ increases from zero (charge neutrality), a sequence of phase transitions occurs. The fully symmetric state undergoes a first-order phase transition at $\mu \approx 0.8W$ in which one of the flavors becomes nearly fully populated (e.g., $\nu_4 \approx 1$), while the densities of all the other flavors are reset to a density a little below the charge neutrality point (see panel (b)). The first-order transition is accompanied by a jump in the total density (panel (c)) and a drop in the inverse compressibility $d\mu/dn$ to zero (panel (d)). After the first-order transition, there is a rapid rise in $d\mu/dn$, diverging when the density of the three partially filled flavors passes through the Dirac point. The inverse compressibility, $d\mu/dn$, then drops gradually, until another first-order transition occurs, where the density of one of the three partially populated flavors jumps up, while the density of the other two flavors jumps down. Note that each of the transitions near $\nu = 1$ and $\nu = 2$ occurs in two nearby first-order steps (the two drops in $d\mu/dn$ near $\nu = 1, 2$ in panel (d)). Finally, between $\nu = 2$ and $\nu = 3$ the two remaining partially filled flavors are split by a second-order transition.

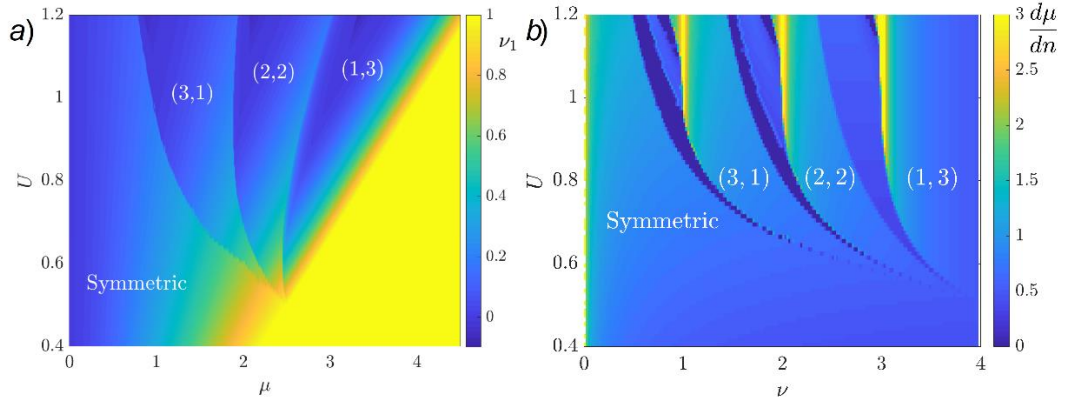

**Figure S10:** a) Phase diagram of the linear DOS model at zero temperature. The color represents the filling factor of the least occupied flavor, $\nu_1$, as a function of chemical potential $\mu$ and the interaction strength $U$, measured in units of $W$. In the symmetric phase, all flavors are populated equally, and spin/valley symmetry is unbroken. ($m_1, m_2$) labels a phase in which $m_1$ flavors are populated less than the other $m_2$. b) Inverse compressibility $d\mu/dn$ as a function of the total filling factor $\nu$ and $U$, in units of the bandwidth $W$.

We find that in the case where the filling of all four bands is of the same sign, it is possible to derive this sequence of phase transitions analytically, and to determine expressions for the phase boundaries. Let us denote by $(m_1, m_2)$ a phase where the first $m_1$ flavors are populated equally, but their fillings are smaller than those of the remaining $m_2$ flavors. We start from the transition from the (4,0) to the (3,1) state. The set of four quadratic equations (10) admit solutions with of the form

$$\nu_\alpha = (W/U)^2 \left[\frac{1}{2} + \frac{s_\alpha(\chi - A_\chi)}{6}\right]^2, \text{ where } \chi = \sum_{\alpha=1}^4 s_\alpha \text{ and } A_\chi = \sqrt{\chi^2 + 12\mu U/W^2 - 15}, \quad (11)$$

which depend only on the dimensionless ratio $\mu U/W^2$. Each distinct solution is characterized by the signs $s_\alpha = \pm 1$. We find three distinct physical solutions up to band permutations. These correspond to the symmetric solution (4,0) or $\{+,+,+,+\}$; and two symmetry broken solutions where the bands are split into (3,1) with $\{+,-,-,-\}$ or (2,2) with $\{+,+,-,-\}$. These solutions can be distinguished by a constant $\chi$, reflecting the sum of $s_\alpha$ in each band. This constant takes the values $\chi = 4$, $\chi = -2$, and $\chi = 0$ for the three aforementioned states, respectively. The total density takes the simple form

$$\nu = (W/U)^2[1 + (A_\chi - \chi)(2A_\chi + \chi)/18], \quad (12)$$

from which we can directly derive the inverse compressibility,

$$d\mu dv = 3UA_\chi\chi - 4A_\chi. \quad (13)$$

The mean field energies of these solutions are shown in figure S12(b). At $\mu U/W^2 = 11/12$ there is a first order phase transition from the symmetric state (4,0) into the (3,1) state. At a larger value of $\mu U/W^2$, a (2,2) solution also appears. Its energy is higher than that of the (3,1) solution. However, the energy difference between the two is relatively small (about $0.1W$ for $U \sim W$). This suggests that small perturbations, such as an applied magnetic field (not considered within our model), can change the balance between different ground states. Indeed, we observe that in the experiment the compressibility near $\nu = 1$ depends strongly on the applied parallel magnetic field.

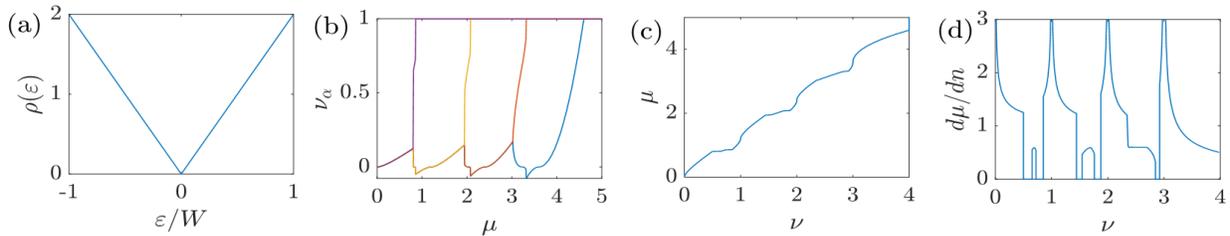

**Figure S11:** Results for linear density of states model with $U = 1.2W$. (a) DOS, (b) $\nu_\alpha$ as a function of $\mu$, (c) $\mu$ as a function of $\nu = \sum_\alpha \nu_\alpha$, (d) $d\mu/dn$ as a function of $\nu$.

The previous solutions did not account for the fact that the linearly dispersive bands have a massive end at $\varepsilon = \pm W$, and that some bands may be fully filled while others are only partially filled. One can similarly address the general case with some number $N_{full}$ of fully occupied bands, and find the locations of the phase boundaries between the different states. We will not provide the full details here.

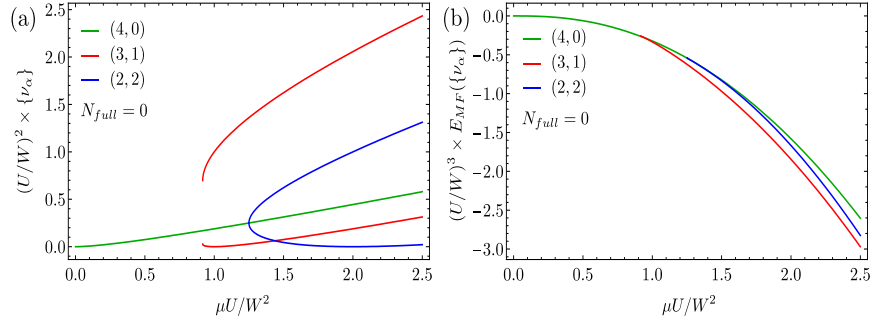

While the precise sequence of the phase transitions as a function of $\mu$ depends on $U$, the overall shape of $d\mu/dn$ is qualitatively the same in the region $11W/12 \lesssim U < U_c = 4W/3$. As we shall see next, the qualitative features of the solution of the linear DOS model remain unchanged even for more complicated (and realistic) models of the DOS.

**Figure S12:** Analytical solutions for the linear density of states model without any fully occupied bands. It describes the states competing close to the first phase transition around $\nu = 1$. We find that in addition to the symmetric state $(4,0)$, there are two possibe symmetry broken solutions with the bands separated in $(3,1)$ or $(2,2)$ groups. In panel (a) we show the filling $\nu_\alpha$ of all bands in each solution; while in panel (b) we show the mean field energy $E_{MF}(\{\nu_\alpha\})$ of these solutions as a function of the single parameter $\mu U/W^2$.

## SI12. Theory: Effect of the Van Hove Singularity

We now consider the effects of van Hove singularities in the narrow bands on the phase diagram and the inverse compressibility. Within the nearly flat band, the DOS must diverge in at least one energy, where the topology of the Fermi surface changes. These van Hove singularities in the DOS have been argued to play a role in stabilizing the correlated insulator phases in TBG[30]. Within our model, we find that in the intermediate coupling regime, $U/W = O(1)$, the sequence of phase transitions as a function of density is not strongly affected by the presence of van Hove (vH) singularities in the DOS near the middle of the band. To exemplify this point, we consider the case of a high-order vH singularity[31], where the DOS diverges even more strongly than in an ordinary vH singularity. In a generic vH point in two spatial dimensions, the DOS diverges logarithmically; in contrast, in a higher-order vH singularity where two vH points meet in a low-symmetry point of the Brillouin zone, $\rho \propto |E|^{-1/4}$. [31] To mimic this situation, we consider the following model DOS:

$$\rho(\varepsilon) = \frac{a|\varepsilon|}{W^2} \Theta(W - |\varepsilon|) \left|\frac{\varepsilon_0}{\varepsilon_v - |\varepsilon|}\right|^{-1/4}. \tag{20}$$

Here, $a$ is a normalization constant chosen such that $\int_0^W d\varepsilon\, \rho(\varepsilon) = 1$. We choose $\varepsilon_v = 0.7W$, corresponding to a high-order vH singularity near $\nu_\alpha = 0.5$ in each flavor, and $\varepsilon_0 = W$. The results for $U = W$, shown in figure S13, are strikingly similar to those of the linear DOS model (figure S11). Qualitatively similar results are found for the case of an ordinary logarithmic vH singularity (not shown).

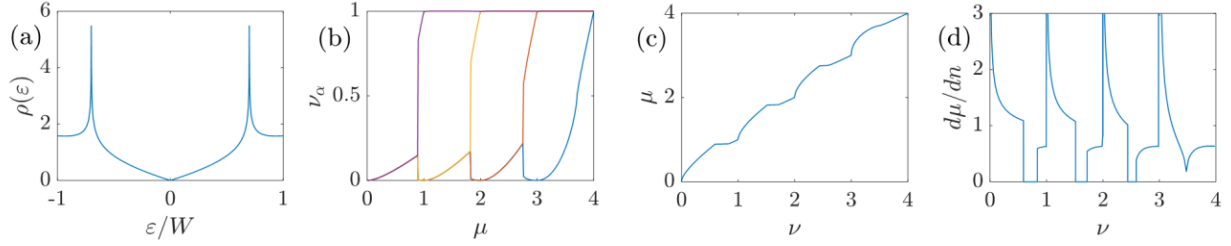

**Figure S13:** Same as Figure S10 for a model DOS with a high-order van Hove singulariy ($\rho(\varepsilon)$ is given by Eq. 20). Here, $U = W$, $\varepsilon_0 = W$, $\varepsilon_v = 0.7W$.

At weaker values of the coupling, the model (20) behaves qualitatively differently from the linear DOS model. In particular, since the DOS diverges at the vH points, there is always an instability towards a spin/valley symmetry broken state, even for arbitrarily small $U$, when the Fermi energy is close to $\varepsilon_v$. This is because the Stoner criterion $\rho(\varepsilon_F)U = 1$ is always satisfied when the Fermi energy $\varepsilon_F$ is sufficiently close to the vH singularity. At intermediate to strong values of the coupling, however, the effects of the vH singularity are much less pronounced, since the phase transitions occur generically when the density is far from the vH density, and the transitions are mostly first order, involving large changes in the populations in the individual flavors (the change in the overall density through the transitions remains small). We therefore conclude that the essential features of the DOS needed to reproduce the qualitative shape of the experimentally measured $d\mu/dn$ are the suppression of $\rho(\varepsilon)$ near charge neutrality due to the Dirac points, and the discontinuity of the DOS near the top and bottom edges of the flat bands, consistent with a quadratic dispersion in two dimensions.

It is worth noting that in figure S13, the last transition before the band gets completely filled (at $\mu \approx 2.8W$) is discontinuous, unlike in the linear DOS model where this transition is of second order. Also, the sharp dip in $d\mu/dn$ at $\nu \approx 3.5$ (figure S13d) is due to crossing the vH singularity.

In addition, we show in figure S14 the results of a mean-field calculation using as an input the DOS of the continuum Bistritzer-MacDonald model near the magic angle, with an interaction strength comparable to the bandwidth. The results are qualitatively similar to those of the linear DOS model, although the asymmetric features in $d\mu/dn$ are much less pronounced then observed in the experiment.

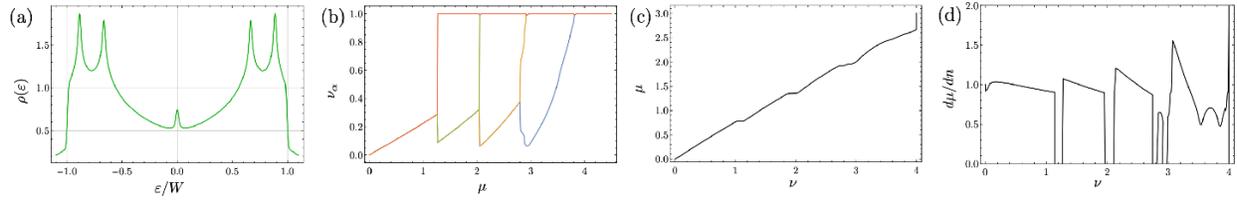

**Figure S14:** Same as Figure S10 for with the DOS obtained for the continuum model of Ref.[18] with the single parameter $\alpha = 0.612$ corresponding to an angle of $\theta \sim 1.04$ degrees. We use $U/W = 1.1$, where the bandwidth $W$ is defined by the sharp drop in the density of states.

## SI13. Theory: Effects of asymmetry in the conduction and valence flat bands

The experimentally measured $d\mu/dn$ exhibits sawtooth-like features around integer filling factors. Within our model, this feature is reproduced if one assumes that the single-particle DOS near the CNP is much lower than that near the top of the conduction and bottom of the valence bands. This difference is directly reflected in the compressibility before the integer fillings: when the density approaches an integer filling factor from below, the DOS at the Fermi level is dominated by the nearly filled flavors with a high DOS, whereas above the integer filling the DOS at $E_F$ is dominated by the flavors whose density is reset to near the CNP, where the DOS is minimal.

To demonstrate the relation between the asymmetry of $d\mu/dn$ around the integer fillings and the DOS of the single-particle bands, we use the following model DOS:

$$\rho(\varepsilon) = \frac{2|\varepsilon|}{W\varepsilon_0}\Theta(\varepsilon_0 - |\varepsilon|) + \frac{2(W-|\varepsilon|)}{W(W-|\varepsilon_0|)}\Theta(|\varepsilon| - \varepsilon_0)\Theta(W - |\varepsilon|). \tag{21}$$

This form describes a triangular-shaped $\rho(\varepsilon)$ with a maximum at $\varepsilon = \varepsilon_0$. It reduced to Eq. (8) in the limit $\varepsilon_0 \to W$. Figure S15 shows the calculated $d\mu/dn$ corresponding to $\varepsilon_0 = 0.3W$, $0.5W$, and $0.7W$. The trend is clearly visible: for $\varepsilon_0 = 0.5W$, the peak in $d\mu/dn$ at $\nu = 2$ is symmetric. For $\varepsilon_0 = 0.7W$, the slope in $d\mu/dn$ is larger at densities below the peak, whereas for $\varepsilon_0 = 0.3W$ the asymmetry of the peak is opposite. The peaks near the other integer fillings follow a similar qualitative dependence on the asymmetry of the single-particle DOS.

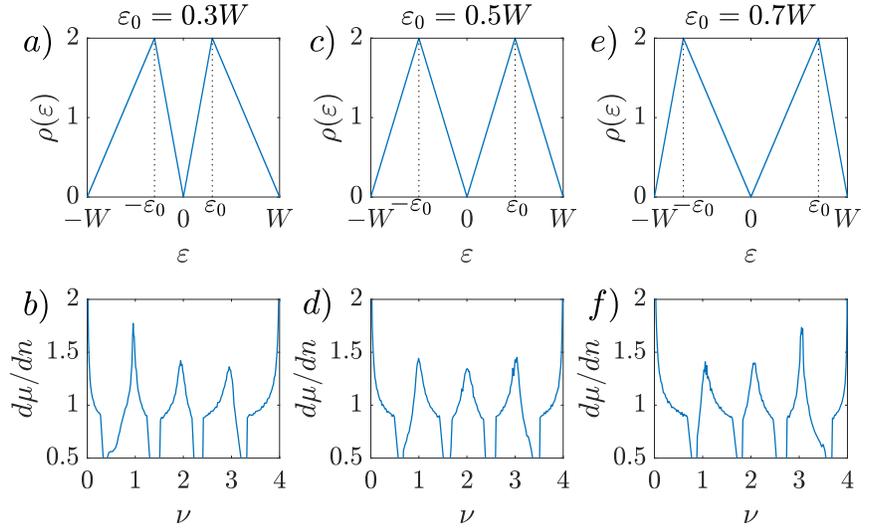

**Figure S15:** Effect of the asymmetry in the single-particle DOS on $d\mu/dn$. The DOS in these calculations is given by Eq. (21), with $\varepsilon_0 = 0.3W$ (panels a,b), $\varepsilon_0 = 0.5W$ (panels c,d), and $\varepsilon_0 = 0.7W$ (panels e,f). Here, $U = 0.85W$.

## SI14. Theory: Finite Temperature

It is straightforward to extend the Hartree-Fock calculation to finite temperature. This is done by minimizing the variational free energy of a Gibbs distribution generated by $H_{MF}$ in Eq. (2) with respect to $\mu_\alpha$. The variational grand potential at temperature $T$ is given by

$$\frac{\Phi_{MF}(\{\mu_\alpha\})}{N} = \sum_\alpha f(\mu_\alpha + \mu) + \frac{U}{2}\sum_{\alpha\neq\beta} \nu(\mu_\alpha + \mu)\nu(\mu_\beta + \mu) + \sum_\alpha \mu_\alpha \nu(\mu_\alpha + \mu) \tag{22}$$

$$\text{where} \quad f(\mu) = -T\int_{-\infty}^{\infty} d\varepsilon\, \rho(\varepsilon)\left[\log\left(1 + e^{-\frac{\varepsilon-\mu}{T}}\right) + \frac{\varepsilon-\mu}{T}\Theta(-\varepsilon)\right], \tag{23}$$

and $\quad \nu(\mu) = \int_{-\infty}^{\infty} d\varepsilon \rho(\varepsilon) \left(\frac{1}{1+e^{(\varepsilon-\mu)/T}} - \Theta(-\varepsilon)\right).$ (24)

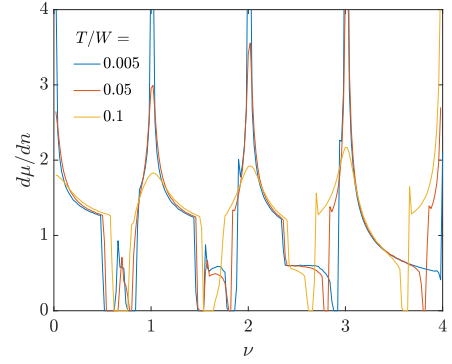

Figure S16 shows the inverse compressibility as a function of $\nu$ for different temperatures in the linear DOS model with $U = 1.2W$. As long as the temperature is sufficiently small compared to $W$, its main effect is the smearing of the sharp asymmetric features in $d\mu/dn$, accompanied by small shifts of the location of the dips that precede these features, consistent with the experimental obervations.

**Figure S16:** $d\mu/dn$ as a a function of the filling factor $\nu$ for different temperatures for the linear DOS model with $U = 1.2W$.

## SI15. Theory: Effects of Long Range Interactions

The long range of the Coulomb interaction between the electrons has several important effects on the experimental measurements and the theoretical considerations that we report. First, it is this long ranged interaction that enables the use of our method to measure the compressibility, a method based on measuring the electric field generated by charged plates at long distances. Second, this long range also requires care in the identification of the quantity that we measure, through the distinction between capacitance, geometric capacitance and quantum capacitance. And third, the long range interaction affects the way the system behaves in the phase separation regime that accompanies first order phase transitions. We now comment on these issues.

Our system may be modeled as composed of two parallel two-dimensional plates separated by a distance $d$, in which one plate is a twisted bi-layer graphene system, and the other is a metallic plate, the back-gate. We assume the metal to have an infinite compressibility, and an instantaneous response. Under these conditions, the metallic plate introduces image charges to the charges in the TBG system, modifying the Coulomb interaction between electrons in the TBG to be

$$V(r) = \frac{e^2}{\varepsilon}\left(\frac{1}{r} - \frac{1}{\sqrt{r^2+(2d)^2}}\right).$$ (25)

Here, $\varepsilon$ is the dielectric constant of the insulator between the two plates. This effective interaction crosses over from the usual Coulomb form $V(r) \sim e^2/\varepsilon r$ at short distances, $r \ll 2d$, to a dipolar form, $V(r) \sim 2e^2d^2/\varepsilon r^3$ at long distances.

Within the canonical ensemble, and under the assumption that the two-plates system as a whole is charge neutral, its energy is

$$E_{tot} = \frac{Q^2}{2C_g} + \langle H_{kin}\rangle + \frac{1}{2}\sum_{q\neq 0} \tilde{V}(q)\langle \rho_q \rho_{-q}\rangle,$$ (26)

where the first term is the electrostatic energy carried by the electric field between the plates ($C_g = \varepsilon A/4\pi d$ is the geometric capacitance, where $A$ is the system's area, and $Q$ is the total charge in the TBG), while the second and third terms give the energy of the TBG plate. In the second term $H_{kin}$ is the single-particle part of the Hamiltonian, while in the third $\tilde{V}$ is the Fourier transform of the potential (25) and $\rho_q$ is the density operator. The $q = 0$ part of the interaction is the first term. The second derivative $\frac{\partial^2 E_{tot}}{\partial Q^2}$ is the inverse of a thermodynamic susceptibility, which must be positive for the system to be stable. It is equal to the inverse of the total capacitance, $C^{-1}$, of the capacitor formed by the two plates. It is useful to define the *quantum capacitance $C_q^{-1}$* through

$$C^{-1} = C_g^{-1} + C_q^{-1}. \tag{27}$$

The inverse quantum capacitance is then the second derivative of the last two terms in (26) with respect to the total charge $Q$.

The method that we employ for our measurements is based on the introduction of a third metallic plate[32] (in our experiment, this is the island of the SET), and measuring the electrostatic potential that it experiences (see Sec. S2 for the equivalent circuit and further details). When the perturbation resulting from this third plate is small, the potential it experiences is proportional to $C_q^{-1}$. By itself, thermodynamical stability does not impose any restrictions on the sign of $C_q^{-1}$.

We now turn to the effect of the long range Coulomb interaction on the charge distribution when the system is close to a first order phase transition, and the way it is reflected in $C^{-1}$ and $C_q^{-1}$. The presence of the back-gate, as reflected in the second term of (25), makes the total energy of the system extensive. Were the interaction between the electrons short-ranged, a first order transition would follow the picture described by the Maxwell construction (figure S17a). With the total density in the system being the controlled parameter, the system would break up into macroscopic domains of the two phases occupying fractions $x$ and $1 - x$ of the system, such that the total density $n$ follows $n = (1 - x)n_A + xn_B$ (here $n_A, n_B$ are the densities of the two phases in the coexistence region determined by equating their chemical potentials, and we assume $n_B > n_A$, such that $x$ grows as $n$ is increased), and the system's energy density $\epsilon_{tot}$ follows

$$\epsilon_{tot}(x) = (1 - x)\epsilon_A + x\epsilon_B. \tag{28}$$

Here $\epsilon_A, \epsilon_B$ are the energy densities of the two phases in the coexistence region. The energy cost associated with interfaces between domains is sub-extensive, and as a consequence the chemical potential $\mu = \frac{\partial \epsilon_{tot}}{\partial n}$ would vary linearly with $n$, making the inverse compressibility $d\mu/dn$ zero in the coexistence region.

The long range Coulomb interaction disfavors the formation of macroscopic domains of different charge densities, due to the charging energy cost. Instead, in the coexistence region the system tends to break up into domains of mesoscopic size[33–35]. The formation of mesoscopic domains makes the interface regions occupy a significant part of the system, and thus adds an extensive interface energy term to the two terms in (28). Remarkably, this term may be negative. We now explore the consequence of such a negative term.

The negative energy contribution originating from interface regions between the two phases vanishes when $x = 0$ or $x = 1$, and is presumably maximized when $x \approx 0.5$. Due to the metallic gate, the interaction decreases as $1/r^3$ at long distances, and therefore the energy of the two phases, as well as that of a macroscopically phase separated state, are extensive. These states are therefore all possible in principle. However, due to the negative energy of the interfaces between the two phases, breaking the system into mesoscopic domains may lower the energy, and thus modifies $\epsilon_{tot}(n)$ in the coexistence region. Thermodynamical stability requires that for any value of $x$ the energy has a positive curvature, and forbids a cusp in which the slope suffers a discontinuous downturn. As a consequence of these requirements, the introduction of mesoscopic domains expands the density range of the coexistence region relative to the range expected from a naive Maxwell construction. Schematically, then, the line $\epsilon_{tot}(x)$ is modified to follow the curve in figure S17b.

This modification affects, of course, the second derivative of the energy with respect to the density, $C^{-1}$. As can be inferred from figure S17b, $C^{-1}$ decreases upon entering the coexistence region, although it is generally non-zero. Since the geometric capacitance is independent of density, the measured $C_q^{-1}$ is reduced in the coexistence region as well.

These general considerations imply that the coexistence region is characterized by a smaller value of $C_q^{-1}$ than in the two nearby phases. Unfortunately, the precise magnitude of $C_q^{-1}$ in the coexistence region depends on details, and is not easy to predict. In particular, it depends on the length scale $\lambda(x)$ of the domains in the coexistence region, which in turns depends sensitively on microscopic parameters[33–35]. Beginning from the limit where long-range interactions are very weak, $\lambda$ diverges, and the results for short-range interactions are recovered; in particular, in the coexistence region, $C^{-1}$ approaches 0, and hence $C_q^{-1}$ approaches $-C_g^{-1}$. As the strength of the long-range interactions increases, $C_q^{-1}$ increases, and can be of either sign. We leave a full theoretical analysis of this problem to future work.

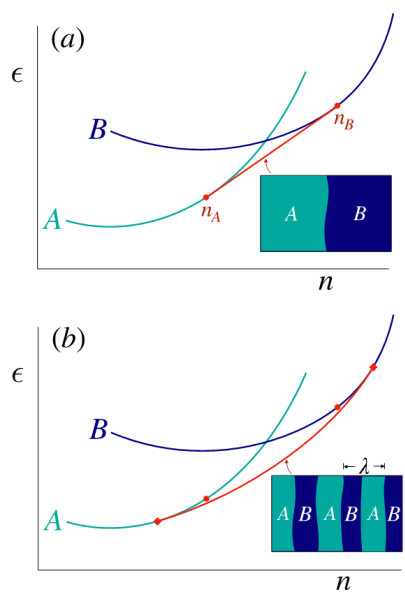

**Figure S17:** (a) Schematic depiction of the energy density $\epsilon$ as a function of density $n$ near a first-order transition for a system with short-range interactions. The red line is determined by a Maxwell construction. At its two ends, it is tangent to the $\epsilon(n)$ curves of the two phases. In this region, the system phase-separates into macroscopic domains of the two phases $A$ and $B$. (b) Energy diagram for a system with long-range (Coulomb) interactions screened by a metallic gate. In this case, the coexistence region is characterized by mesoscopic domains of the two phases. The characteristic size $\lambda$ of the domains is determined by a balance between energy gain associated with phase separation and the charging energy that favors local charge neutrality. The coexistence region, marked by a red curve, is larger than the region of the Maxwell construction (whose ends are indicated by red circles).

## SI16. The Predicted Landau Fans from the Dirac Revivals Model.

The model of Dirac revivals gives specific predictions regarding the expected quantum oscillations as a function of band filling. These are shown in fig. S18, where in panel a we reproduced fig. 4d from the main text, which shows the partial flavor occupation as function of filling factor, and panel b illustrates the predicted Landau fans. Visibly, the bands start from the CNP with 4-fold flavor degeneracy. As $\nu = 1$ is approached, one band takes all the carriers resetting the other three bands back to near the CNP. Consequently, just after $\nu = 1$ the flavor degeneracy reduces to 3-fold. The process repeats at $\nu = 2$ and 3, leading to 2-fold and 1-fold degeneracies just after these filling factors. This is reflected in the Landau fans in the bottom panel. This 4,3,2,1-fold predicted flavor degeneracy is indeed the one observed experimentally in global transport[12–15] and local scanning squid[36] measurements, with one possible exception, above $\nu = 1$, where transport data were interpreted to show 2-fold degeneracy whereas the Dirac revivals model predicts 3-fold degeneracy. We should note, though, that in this region the data is considerably sparser than near other filling factors and therefore can be less conclusive. In fact, scanning squid measurements also find above $\nu = 1$ a periodicity that is harder to simply interpret as 2-fold.

Another important feature of the Landau fan explained by the Dirac revivals model is their direction in the $\nu - B$ plane. Transport and scanning squid measurements have consistently observed that the Landau fans that emerge from integer fillings appear only on one side of the integer filling, pointing *away* from the CNP. This observation holds for Landau levels both in the valence and the conduction flat bands. There is only one exception for this rule, where near $\nu = \pm 4$ (full bands), Landau fans that point *toward* the CNP are occasionally observed[8].

These observations can be naturally explained by the Dirac revivals model. The directionality of the Landau fans depends on the polarity of the carriers. At the bottom of the conduction flat band (just above the CNP) the carriers have electronic character, but near its top they have a hole character. The fermi surface transitions between these two polarities at a van Hove Singularity (VHS) that is generally believed to be near the center of the band. Within a simple single particle band filling, one would therefore expect the Landau fans to switch their direction somewhere around $\nu = 2$. Such mirror symmetry around $\nu = 2$ should also be generically true in the presence of interactions, as long as electron-hole symmetry around the center of the conduction band is valid. The Dirac revivals model suggest a different scenario: here, after every integer $\nu$ the bands start refilling from the bottom of the conduction band (the CNP), but never make it to the VHS, because they are always reset back before reaching the VHS. Thus, the carriers always retain an electronic character (red shaded region in fig. S18a), explaining why the Landau fans always point away from the CNP for $\nu = 0 - 3$. In contrast, the last band to fill, between $\nu = 3$ to 4, continuously fills from its bottom to its top and therefore has to cross the VHS at a certain point. Thus it starts with an electron character and ends with a hole character. Consequently, the Landau fans emerging from $\nu = 3$ will point away from the CNP, whereas the ones emerging from $\nu = 4$ should point toward the CNP. In the valance flat band our model predicts the same behavior but mirror reflected around the CNP. We note that while our simple model predicts that the hole-like Landau fan emerging from $\nu = 4$ should have 1-fold degeneracy, experiments see a 4-fold hole-like fan, a discrepancy that highlights that close to full band the flavor symmetry breaking should disappear. Overall, however, we find that the Dirac revivals model captures the main features of the observed Landau fans rather well.

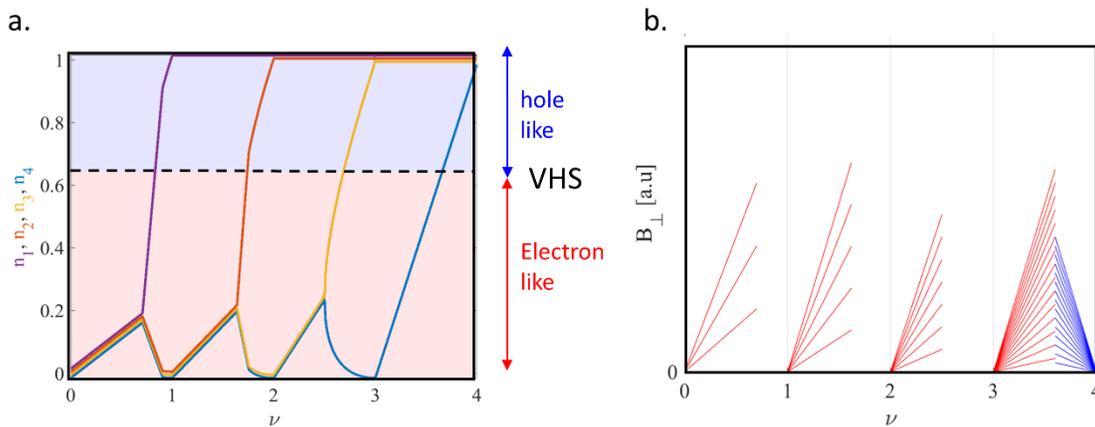

**Figure S18: Predicted landau fan degeneracies and polarities from the Dirac revival model. a.** Partial flavor population, $n_i$, as a function of $\nu$. The figure is reproduced from fig. 4d in the main text, and we have marked by red and blue the regions that should have electron and hole like carriers, separated by the position of the van Hove Singularity, which is generally believed to be around the center of the band. **b.** Landau fans predicted from the Dirac revivals model, with a 4,3,2,1-fold degeneracy after $\nu = 0,1,2,3$. Red/blue corresponds to electron/hole-like character shown in panel a.